\documentclass{emulateapj}
\usepackage{apjfonts}
\usepackage{natbib}
\usepackage{graphics}
\usepackage{multirow}
\usepackage{amsbsy}
\usepackage{mathrsfs}
\usepackage{footmisc}
\usepackage{hyperref}
\usepackage{amsmath}
\usepackage{verbatim}
\hypersetup{
  colorlinks = true,
  urlcolor = black,
  linkcolor=black,
  citecolor=black
}

\def\bicep{{\sc Bicep}}

\def\dasi{{\sc Dasi}}

\def\bicepone{{\sc Bicep1}}
\def\biceptwo{{\sc Bicep2}}

\def\keckarray{{\it Keck Array}}
\def\planck{{\it Planck}}
\def\quiet{{\sc Quiet}}

\def\QUAD{{\sc QUaD}}
\def\cbi{{\sc CBI}}
\def\capmap{{\sc CAPMAP}}
\def\maxipol{{\sc MAXIPOL}}

\def\boom{{\sc Boomerang}}
\def\wmap{WMAP}

\def\polarbear{{\sc Polarbear}}

\def\actpol{ACTPol}
\def\sptpol{SPTpol}

\def\synfast{{\tt synfast}}
\def\healpix{{\tt Healpix}}

\def\master{{\tt MASTER}}

\def\gcp{{\tt gcp}}

\newcommand{\ukcmbrts}{ $\mu\mathrm{K}_{\mathrm{\mbox{\tiny\sc cmb}}}\sqrt{\mathrm{s}}$ }

\def\krjtxt{{{$\mathrm{K}_{\mathrm{\mbox{\tiny\sc rj}}}$}}}
\def\deg{^\circ}
\def\emode{$E$-mode}
\def\bmode{$B$-mode}

\def\lcdm{$\Lambda$CDM}

\newcommand{\upd}{\mathrm{d}}

\def\apj{Astrophys.\ J.}

\shorttitle{Measurements of \bmode\ Polarization at Degree Angular
Scales and 150~GHz by the \keckarray}
\shortauthors{\keckarray\ and \biceptwo\ Collaborations}
\submitted{to be submitted to \apj --- draft version \today}
\begin{document}

\title{\biceptwo\ / \keckarray\ V: Measurements of \bmode\ Polarization at Degree Angular
Scales and 150~GHz by the \keckarray}

\author{\keckarray\ and \biceptwo\ Collaborations: 
P.~A.~R.~Ade\altaffilmark{1}}
\author{Z.~Ahmed\altaffilmark{2}}
\author{R.~W.~Aikin\altaffilmark{3}}
\author{K.~D.~Alexander\altaffilmark{4}}
\author{D.~Barkats\altaffilmark{5}}
\author{S.~J.~Benton\altaffilmark{6}}
\author{C.~A.~Bischoff\altaffilmark{4}}
\author{J.~J.~Bock\altaffilmark{3,7}}
\author{J.~A.~Brevik\altaffilmark{3}}
\author{I.~Buder\altaffilmark{4}}
\author{E.~Bullock\altaffilmark{8}}
\author{V.~Buza\altaffilmark{4}}
\author{J.~Connors\altaffilmark{4}}
\author{B.~P.~Crill\altaffilmark{3,7}}
\author{C.~D.~Dowell\altaffilmark{7}}
\author{C.~Dvorkin\altaffilmark{4}}
\author{L.~Duband\altaffilmark{9}}
\author{J.~P.~Filippini\altaffilmark{3,10}}
\author{S.~Fliescher\altaffilmark{11}}
\author{S.~R.~Golwala\altaffilmark{3}}
\author{M.~Halpern\altaffilmark{12}}
\author{M.~Hasselfield\altaffilmark{12}}
\author{S.~R.~Hildebrandt\altaffilmark{3,7}}
\author{G.~C.~Hilton\altaffilmark{13}}
\author{V.~V.~Hristov\altaffilmark{3}}
\author{H.~Hui\altaffilmark{3}}
\author{K.~D.~Irwin\altaffilmark{2,13,14}}
\author{K.~S.~Karkare\altaffilmark{4}}
\author{J.~P.~Kaufman\altaffilmark{15}}
\author{B.~G.~Keating\altaffilmark{15}}
\author{S.~Kefeli\altaffilmark{3}}
\author{S.~A.~Kernasovskiy\altaffilmark{2,X}}
\author{J.~M.~Kovac\altaffilmark{4}}
\author{C.~L.~Kuo\altaffilmark{2,14}}
\author{E.~M.~Leitch\altaffilmark{16,17}}
\author{M.~Lueker\altaffilmark{3}}
\author{P.~Mason\altaffilmark{3}}
\author{K.~G.~Megerian\altaffilmark{7}}
\author{C.~B.~Netterfield\altaffilmark{6,18}}
\author{H.~T.~Nguyen\altaffilmark{7}}
\author{R.~O'Brient\altaffilmark{7}}
\author{R.~W.~Ogburn~IV\altaffilmark{2,14}}
\author{A.~Orlando\altaffilmark{15}}
\author{C.~Pryke\altaffilmark{11,8}}
\author{C.~D.~Reintsema\altaffilmark{13}}
\author{S.~Richter\altaffilmark{4}}
\author{R.~Schwarz\altaffilmark{11}}
\author{C.~D.~Sheehy\altaffilmark{16,17}}
\author{Z.~K.~Staniszewski\altaffilmark{3,7}}
\author{R.~V.~Sudiwala\altaffilmark{1}}
\author{G.~P.~Teply\altaffilmark{3}}
\author{K.~L.~Thompson\altaffilmark{2}}
\author{J.~E.~Tolan\altaffilmark{2}}
\author{A.~D.~Turner\altaffilmark{7}}
\author{A.~G.~Vieregg\altaffilmark{16,17}}
\author{A.~C.~Weber\altaffilmark{7}}
\author{J.~Willmert\altaffilmark{11}}
\author{C.~L.~Wong\altaffilmark{4}}
\author{K.~W.~Yoon\altaffilmark{2,14}}

\altaffiltext{1}{School of Physics and Astronomy, Cardiff University, Cardiff, CF24 3AA, UK}
\altaffiltext{2}{Department of Physics, Stanford University, Stanford, CA 94305, USA}
\altaffiltext{3}{Department of Physics, California Institute of Technology, Pasadena, CA 91125, USA}
\altaffiltext{4}{Harvard-Smithsonian Center for Astrophysics, 60 Garden Street MS 42, Cambridge, MA 02138, USA}
\altaffiltext{5}{Joint ALMA Observatory, ESO, Santiago, Chile}
\altaffiltext{6}{Department of Physics, University of Toronto, Toronto, ON, Canada}
\altaffiltext{7}{Jet Propulsion Laboratory, Pasadena, CA 91109, USA}
\altaffiltext{8}{Minnesota Institute for Astrophysics, University of Minnesota, Minneapolis, MN 55455, USA}
\altaffiltext{9}{SBT, Commissariat \`a l'Energie Atomique, Grenoble, France}
\altaffiltext{10}{Department of Physics, University of Illinois at Urbana-Champaign, Urbana, IL 61820 USA}
\altaffiltext{11}{Department of Physics, University of Minnesota, Minneapolis, MN 55455, USA}
\altaffiltext{12}{Department of Physics and Astronomy, University of British Columbia, Vancouver, BC, Canada}
\altaffiltext{13}{National Institute of Standards and Technology, Boulder, CO 80305, USA}
\altaffiltext{14}{Kavli Institute for Particle Astrophysics and Cosmology, SLAC National Accelerator Laboratory, 2575 Sand Hill Rd, Menlo Park, CA 94025, USA}
\altaffiltext{15}{Department of Physics, University of California at San Diego, La Jolla, CA 92093, USA}
\altaffiltext{16}{Kavli Institute for Cosmological Physics, University of Chicago, Chicago, IL 60637, USA}
\altaffiltext{17}{Department of Physics, Enrico Fermi Institute, University of Chicago, Chicago, IL 60637, USA}
\altaffiltext{18}{Canadian Institute for Advanced Research, Toronto, ON, Canada}
\altaffiltext{X}{Corresponding author: sstokes@stanford.edu}

\begin{abstract}
The \keckarray\ is a system of cosmic microwave background (CMB)
polarimeters, each similar to the \biceptwo\ experiment.
In this paper we report results from the 2012 and 2013 observing seasons,
during which the \keckarray\ consisted of five receivers
all operating in the same (150~GHz) frequency band and observing
field as \biceptwo.
We again find an excess of \bmode\ power over the lensed-\lcdm\ expectation
of $>5\sigma$ in the range $30<\ell<150$ and confirm that this
is not due to systematics using jackknife tests and simulations
based on detailed calibration measurements.
In map difference and spectral difference tests these new data
are shown to be consistent with \biceptwo.
Finally, we combine the maps from the two experiments
to produce final $Q$ and $U$ maps which have a depth of 57~nK$\,$deg
(3.4~~$\mu$K$\,$arcmin)
over an effective area of 400 deg$^2$ for an
equivalent survey weight of 250,000~$\mu$K$^{-2}$.
The final $BB$ band powers have noise uncertainty a factor of 2.3
times better than the previous results, and a significance
of detection of excess power of $>6\sigma$.
\end{abstract}


\section{Introduction}

Precision polarimetry of the cosmic microwave background (CMB)
has become a mainstay of observational cosmology.  The \lcdm\ model
predicts a polarization of the CMB at the level of a few $\mu$K,
with a characteristic \emode\ pattern.  The $EE$ power spectrum has
been detected over a wide range of angular scales by many experiments,
including
\dasi~\citep{kovac02},
\capmap~\citep{barkats04,bischoff08}, 
\cbi~\citep{readhead04,sievers07}, 
\boom03~\citep{montroy06}, 
\wmap~\citep{page07,bennett13}, 
\maxipol~\citep{wu07}, 
\QUAD~\citep{pryke09,brown09}, %
\bicepone~\citep{chiang10,barkats14},
\quiet~\citep{quiet11,quiet12},
\polarbear~\citep{polarbear14},
\biceptwo~\citep{b2respap14},
\actpol~\citep{naess14},
\sptpol~\citep{crites14},
and \planck~\citep{planckI}.
These measurements have been in broad agreement with theoretical
expectations and other cosmological data sets.
Improved $EE$ power spectrum data
are important because they may eventually constrain the \lcdm\ model parameters
better than cosmic variance limited CMB temperature data~\citep{rocha04,galli14}.

Of greater interest is the \bmode\ component of the polarization pattern.
Though the $EE$ power spectrum is higher, the $BB$ power spectrum
is more sensitive to new physics because the linear density perturbations
at the surface of last scattering, which are the main source of $TT$ and $EE$ power,
cannot generate \bmode\ power.
On small angular scales, $BB$ power instead
arises from the gravitational lensing of \emode\ power by the large scale
structure of the universe~\citep{zaldarriaga98}.  The lensing $BB$ power
thus cleanly traces the growth of structure, complementary to other methods,
providing information about possible extensions to \lcdm\ such as neutrino mass or
a nontrivial dark energy equation of state.
Measurements by
\sptpol~\citep{hanson13}
\polarbear~\citep{polarbear6645,polarbear6646,polarbear14},
and \biceptwo~\citep{b2respap14}
have provided the first evidence of $BB$ power from gravitational lensing.

On large angular scales, lensing contributes only a small amount of $BB$ power.
However, inflationary gravitational waves (IGW) may be a source of
$BB$ power on these scales~\citep{polnarev85,seljak97a,kamionkowski97,seljak97b}.
The recent detection by \biceptwo\ of \bmode\ power on degree angular scales
in excess above the lensing expectation is especially
exciting because it could be evidence of primordial gravitational waves
and cosmic inflation~\citep{b2respap14}.
The contribution of foregrounds to the observed \biceptwo\ signal is uncertain,
and preliminary data from \planck\ have suggested that polarized dust in the
\biceptwo\ field may be brighter than models had predicted~\citep{planckiXXX}.
Regardless, a confirmation of the \biceptwo\ signal, whether cosmological or galactic
in origin, is a top priority of observational cosmology today~\citep{dodelson14,caligiuri14}.

The \keckarray\ telescope is a microwave polarimeter at the South Pole designed to
follow up the \biceptwo\ observations.  The \keckarray\ quickly deployed
a large number of detectors at 150~GHz by installing five receivers of
a design very similar to \biceptwo\ with minimal changes.
All five receivers were installed in time for the 2012 observing season
and continued, with upgrades, to observe at 150~GHz through the end of 2013.
The modular, multi-receiver structure of the \keckarray\ also allows
individual receivers to be tuned to different frequencies.
Two of the \keckarray\ receivers began observing at 95~GHz in 2014,
which will help to discriminate the signal from foregrounds.

In this paper, we present the results of the 150~GHz observations
by the \keckarray\ of the \biceptwo\ field during the 2012 and 2013 seasons.
We begin with sections describing the \keckarray\ instrument,
calibrations, and analysis methods.
We proceed with the maps and angular power spectra obtained
from this data set and perform extensive internal consistency checks.
The \keckarray\ confirms the \biceptwo\ \bmode\ signal at $>5\sigma$.
We then proceed to test for consistency between the \biceptwo\
and \keckarray\ data, and finally combine the two sets of
maps to a final result.

This paper is the latest in a series of publications by the
\biceptwo\ and \keckarray\ collaborations.
\cite{b2respap14} (hereafter the \biceptwo\ Results Paper)
is directly analogous to this paper.
\cite{b2instpap14} (hereafter the \biceptwo\ Instrument Paper)
presented the full details of the \biceptwo\ instrument---differences
are summarized in~\S\ref{sec:instrument} of this paper.
\cite{b2systpap14} (hereafter the Systematics Paper) presents a detailed
analysis of instrumental systematics, which are treated for the \keckarray\ 2012--13 data
in~\S\ref{sec:systematics} of the current paper.
Two additional papers, \cite{b2beams14} (the Beams Paper) and
\cite{dets2014} (the Detectors Paper), describe the beam characterization
and the detectors for both \biceptwo\ and the \keckarray.

\section{The \keckarray\ instrument}
\label{sec:instrument}

\begin{figure*}
\resizebox{\textwidth}{!}{\includegraphics{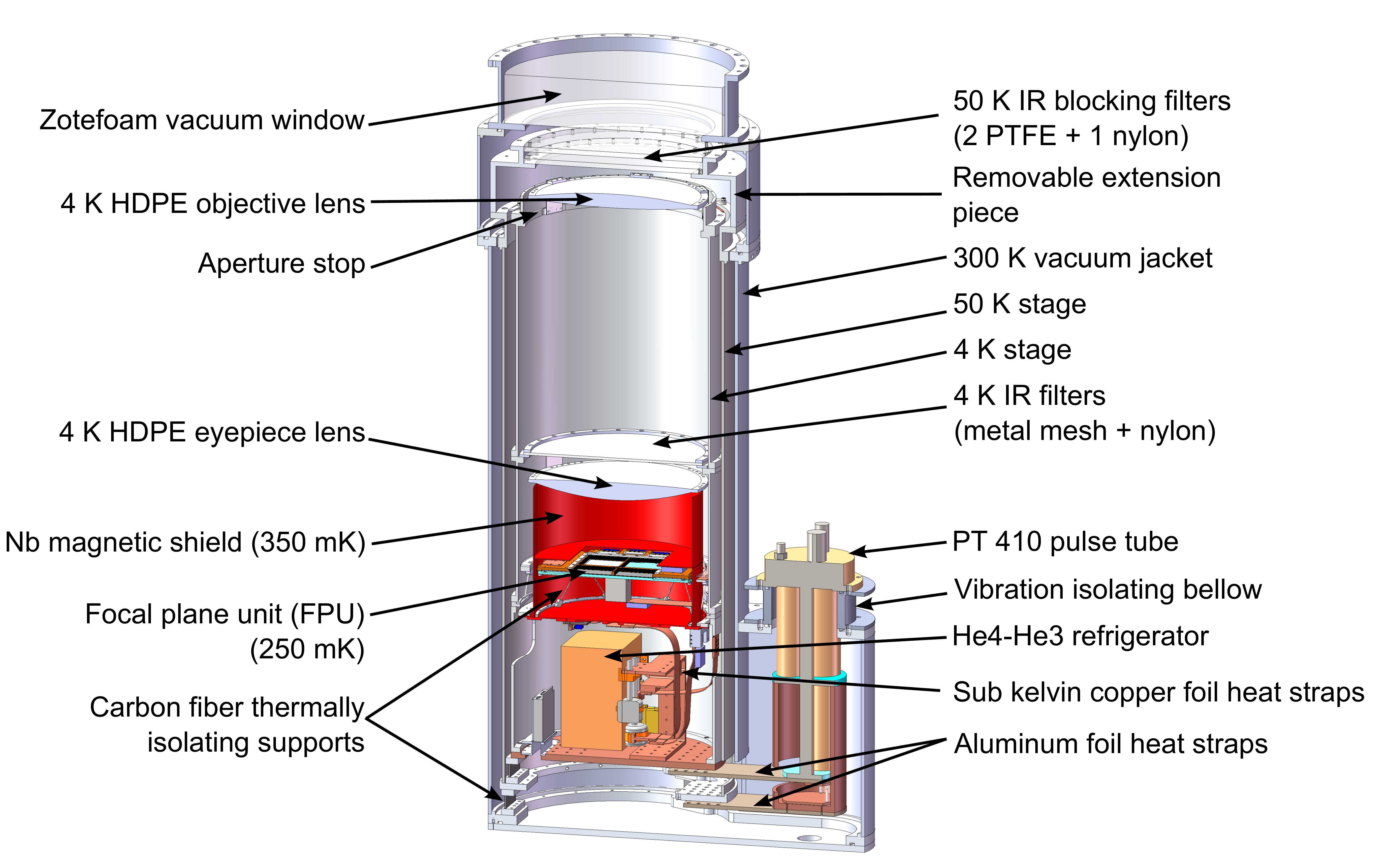}}
\caption{Individual receiver of the \keckarray.  Each receiver
is cryogenic, with a pulse tube refrigerator cooling the
optics to 4~K and a three-stage sorption refrigerator cooling
the focal plane to 270~mK.  The \keckarray\ consists of
five identical receivers on a single telescope mount at the
South Pole.}
\label{fig:keckinstr}
\end{figure*}

The \keckarray\ instrument shares much of its design with \biceptwo,
details of which are presented in the \biceptwo\ Instrument Paper.
In this section, we describe the main features common to both instruments
and the substantive changes and upgrades unique to the \keckarray.
Figure~\ref{fig:keckinstr} shows the receiver design
for the \keckarray.

\subsection{Cryostat and cryogenic system}
\label{sec:cryo}

The \keckarray\ comprises five independent cryostats~\citep{sheehy10}
built by
Atlas Technologies\footnote{\url{http://www.atlasuhv.com/}}.
Inside each cryostat is a closed-cycle,
three-stage ($^4$He/$^3$He/$^3$He) sorption refrigerator~\citep{duband99}
that cools the focal plane unit (FPU) to approximately 270~mK.
Other optical elements are held at cryogenic temperatures to minimize the
thermal load on the FPUs.

The main difference between \keckarray\ and \biceptwo\ is the bulk
refrigeration system.  While \biceptwo\ used a bath of liquid helium,
the \keckarray\ uses a set of Cryomech\footnote{\url{http://www.cryomech.com/}}
PT-410 pulse tube refrigerators.  Each \keckarray\ cryostat
has its own pulse tube refrigerator aligned along the optical axis.
The helium gas is pulsed at a common frequency of 1.2~Hz,
and the pressure in each system is optimized to achieve the lowest base temperature.
After optimization, the pulse tubes'
copper mounting surfaces typically reach 40~K and 3~K with comparable
performance in all five cryostats.
These surfaces are thermally connected
to the telescope insert by stacks of ultra high purity aluminum foil.

\subsection{Optics}
\label{sec:optics}

The \keckarray\ optics use an on-axis, refractive design
which was originally demonstrated in the \bicepone\ telescope~\citep{takahashi10}.
The entire optics chain is essentially unchanged from \biceptwo~\citep{aikin10}.
The two-lens design was chosen to accommodate the flat telecentric
focal plane, with the image of the primary at infinity as
viewed from the focal plane.

The lenses are made from high density polyethylene (HDPE)
and cooled to 4~K.  In order to sufficiently
reduce the infrared loading on the cooling stages,
there is a 3~mm nylon filter and two polytetrafluoroethylene (PTFE)
filters of thickness 12.7~mm and 34.3~mm in the optics path,
cooled to an intermediate stage of 50~K.  A second
nylon filter of 5.2~mm is placed between the objective and
eyepiece lens, heat sunk to 4~K.  A metal mesh low-pass
filter~\citep{ade06} with a cutoff of 8.3~cm$^{-1}$ (225~GHz) is
placed above the final nylon filter to prevent any stray
radiation that was not absorbed by the plastic filters from
thermalizing in the detectors.

All surfaces surrounding the optical path are blackened
with Eccosorb HR10\footnote{Emerson \& Cuming Microwave Products,
Randolph, MA 02368. Phone: 781-961-9600. Web: \url{http://eccosorb.com/}}
cut in half and epoxied with Stycast 2850 loaded with carbon.  The
Stycast covers the HR10 to prevent particulate shedding
during cryogenic cycling.
The lining is designed such that stray
light terminates on cold surfaces.
Later configurations installed after the 2013 observing season included
baffling on the inside of the telescope tube
to further reduce reflections.

The aperture stop is similarly made from a ring of
1.9~cm thick Eccosorb AN-74, beveled at 40$\deg$ with an
inner diameter of 26.4~cm.  Approximately 20\% of the
total throughput is absorbed by the aperture stop.  The
aperture stop is placed on the lower surface of the objective
lens.

\subsection{Focal plane unit}
\label{sec:fpu}

The detectors, developed at Jet Propulsion Laboratory (JPL) for joint use in
the \biceptwo, SPIDER and \keckarray\ experiments,
consist of planar antennas, each an array of slot sub-antennas combined
in phase, and feeding into dual series Ti and Al Transition-Edge Superconducting (TES)
bolometers.  A detailed description is given in the Detectors Paper.
Each pixel consists of two interleaved phased antenna arrays
with orthogonal polarization directions.
The signal is bandpass filtered by a lithographed filter
in stripline after the antenna and terminated on a thermally isolated island
which also contains the series TES.
A single Si tile contains 64 detector pairs and a focal plane unit has 4 tiles.
In each focal plane, 8 detector pairs are left ``dark''.
Dark detectors consist of the complete TES island structure, but are not
connected to their corresponding antennas.

The focal plane units in the \keckarray\ were slightly modified versus those
described in the \biceptwo\ Instrument Paper.  The spacing between the
tiles was increased in order to reduce the electromagnetic coupling between
the pixels near the edge of the tile and the copper plate.
The feed network of the antennas was also redesigned to reduce the dipole beam mismatch
between the polarized pairs, significantly improving the performance versus \biceptwo.

The TES detectors are voltage biased and the current is inductively
coupled to time-domain multiplexing SQUIDs~\citep{dekorte03}.
The \keckarray\ uses the NIST developed MUX09s, which have a gradiometric design
that reduces the sensitivity of the SQUIDs to uniform magnetic fields by
three orders of magnitude in
comparison to the MUX07a design used in \biceptwo~\citep{stiehl11}.

After the deployment of \biceptwo, we discovered that the aliased noise
from the multiplexing system was affecting the overall sensitivity
of the instrument.  One way to mitigate the aliased noise is to
increase the Nyquist inductors that limit the bandwidth of the
detectors.  The choice of inductance is balanced with the need for the
L/R time constant to be fast enough for the detectors to be in stable
negative electrothermal feedback.
The first focal plane produced for the \keckarray\ has Nyquist chips
with an inductance of 1.35~$\mu$H consistent with \biceptwo, and all
subsequent focal planes have an increased inductance of 2~$\mu$H.
This limits the bandwidth to $\le18$~kHz.

\subsection{Readout}
\label{sec:readout}

The configuration of the room temperature electronics that interface with \keckarray\
is similar to that described in the \biceptwo\ Instrument Paper.
A Multi-Channel Electronics (MCE) crate provided by the University of British Columbia
mounts directly to the outside of each cryostat to interface with the SQUIDs,
and supply the detector bias~\citep{mce08}.
The \keckarray\ MCE crates use lower power SQUID series array readout cards
compared to earlier designs in order to stay compatible with
a development program to improve operability on balloon-borne telescopes.

Housekeeping thermometry is read through ``backpacks'' attached to the
cryostats similar to \biceptwo, and the signals are collated and digitized
in a common BLASTbus2 crate provided by the University of Toronto~\citep{benton14}.
Both the housekeeping and detector electronics are connected to a set of Linux-based
computers and recorded to disk using the control software \gcp\ at a sample rate of
20~Hz~\citep{story12}.

\subsection{Mount}
\label{sec:mount}

All five \keckarray\ receivers are attached to a common telescope mount
located at the Martin A. Pomerantz Observatory (MAPO) at the Amundsen-Scott
South Pole Station.
This mount was previously
used for the \dasi~\citep{leitch02} and \QUAD~\citep{hinderks09} experiments.
A new front end ``drum'' for the \keckarray\ cryostats was installed in 2010.
The platform was leveled at that time to account for gradual shifts of the
building on the snow relative to the horizon.

The mount has three axes: elevation, azimuth, and boresight.
The rotation around the boresight is referred to as ``deck rotation''
and allows for cancellation of some systematic effects and/or tests
for their presence.

\section{Characterization}
\label{sec:char}

The \keckarray\ has been extensively characterized in laboratory tests
and with {\it in situ} calibration measurements.  The characterization program
was very similar to the one described in Sections 10 and 11 of the \bicep2
Instrument Paper.  This section summarizes these measurements, focusing
particularly on detector properties that have been reoptimized since the
fabrication of the \bicep2 detectors.  The spectral band, optical efficiency,
and bolometer thermal conductance have a strong effect on the ultimate
sensitivity of the instrument, and are tuned to minimize noise while allowing
stable operation under typical South Pole atmospheric loading conditions.
These detector properties, summarized in Table~\ref{tab:detparams},
are described in Sections~\ref{sec:fts}, \ref{sec:oe}, and \ref{sec:detparam}.
We have also extensively measured the far-field beams with \textit{in situ}
observations of a mast-mounted source.  The beam mapping measurement and
its results are summarized here in Section~\ref{sec:beams} and described
more fully in the Beams Paper.

\begin{table}[t]
\caption{The \keckarray\ Detector Parameters}
\label{tab:detparams}
\begin{center}
\begin{tabular}{lc} 
\hline \hline
\rule[-1ex]{0pt}{3.5ex}  Detector Parameter & Median Value \\
\hline
Optical efficiency, $\eta$ & 24\% \\
Band center, $\langle \nu \rangle$ & 151~GHz \\
Spectral bandwidth, $\Delta\nu$ &42~GHz \\
Normal resistance, $R_N$ & 62 m$\Omega$ \\
Operating resistance, $R_\mathrm{op}$ & 0.68 $R_N$ \\
Saturation power, $P_\mathrm{sat}$ & 9.9~pW \\
Optical loading, $P_\mathrm{opt}$ &3.1~pW \\
Thermal conductance, $G_c$ & 90~pW/K\\
Transition temperature, $T_{c}$ & 520~mK\\
Thermal conductance exponent, $\beta$ & 2.5\\
\hline
\end{tabular}
\end{center}
\end{table}

\subsection{Spectral response}
\label{sec:fts}

The frequency response of the antennas and lumped element filters
was tuned to give a fractional bandwidth of 25\%.
The $\sim 150$~GHz observing band is bracketed by the 118.8~GHz oxygen line on the
low side and the 183.3~GHz water line on the high side.

The frequency response $S(\nu)$ of the 150~GHz detectors was characterized using a
Martin-Puplett Fourier transform spectrometer (FTS)~\citep{karkare14}.
From these spectra, the band
center and the bandwidth are calculated.
The band center is defined to be
\begin{equation}
\label{eqn:ftsbc}
\langle \nu\rangle = \frac{\int \nu I(\nu) S(\nu)\upd \nu}{\int I(\nu) S(\nu) \upd \nu}
\end{equation}
and the bandwidth
\begin{equation}
\label{eqn:ftsbw}
\Delta \nu = \frac{(\int I(\nu) S(\nu)\upd \nu)^2}{\int I^2(\nu) S^2(\nu)\upd \nu},
\end{equation}
where $I(\nu)$ is the source spectrum relative to a Rayleigh-Jeans spectrum.
For a Rayleigh-Jeans spectrum, the 150~GHz detectors were measured to have a band center of
151.5$\pm$1.9~GHz and a bandwidth of 41.8$\pm$1.4~GHz.  The effective band center
shifts to 150.6, 152.8 and 148.6~GHz respectively for a source spectrum of
CMB, dust, and synchrotron radiation using the current best-fit models~\citep{planckiXXX,bennett13}
The standard deviations are dominated by variation from tile to tile,
with smaller variation from detector to detector within a tile.

\subsection{Optical efficiency}
\label{sec:oe}

The end-to-end optical efficiency is defined as the fraction of incident
light absorbed by the detectors.
This is dependent on the losses in the optics, the antennas,
and the bandpass filters.  A higher optical efficiency increases the
sensitivity of the detectors. Because it also increases the optical
loading and photon noise, it must be taken into account when optimizing the
thermal conductivity of the detector.

The optical efficiency is measured in the lab using a beam-filling,
microwave absorbing cone of AN-72.  The power change on the detector for
a source in the Rayleigh-Jeans limit ($h\nu\ll kT$) is:
\begin{equation}
P_\mathrm{opt} = kT\eta \Delta \nu
\end{equation}
where $\eta$ is the optical efficiency and $\Delta \nu$ is the bandwidth
as defined in \S\ref{sec:fts}.  The detector loading was measured with the
cone at both room temperature and liquid nitrogen temperature and
converted into optical efficiency with the measured bandpass of
42~GHz.

The median measured end-to-end optical efficiency of the \keckarray\
receivers was 24\%.
The optical efficiency of the early \keckarray\ detectors
was lower than the 38\% optical efficiency of the \biceptwo\ detectors.
Detector testing after initial \keckarray\ deployment suggested
that the optical efficiency was being reduced by microscopic
stress-induced cracks in the niobium microstrips connecting the antenna
networks to the TES bolometers.  In later generations of \keckarray\ detectors the
Nb film stress was decreased from $\sim 1000$~MPa to $<300$~MPa.
The optical efficiency was observed to increase
as described in more detail in the Detectors Paper.

\subsection{Thermal conductance}
\label{sec:detparam}

As is discussed in the \biceptwo\ Instrument Paper, the
detector parameters can be tuned during fabrication in order
to optimize the noise performance.  In particular, the
thermal conductance can be tuned to minimize the phonon
noise while maintaining a margin of safety ensuring
operability under normal loading conditions.

The phonon noise is the thermal fluctuations from the substrate
to the detector island through the SiN isolation legs.
The noise-equivalent-power (NEP) is
dependent on the thermal conductance $G$ across the legs
(see e.g.~\citet{hiltonirwin2005}) as
  \begin{equation}
		\label{eqn:phonon}
    \mathrm{NEP}=\sqrt{4k_BGT_c^2F}.
  \end{equation}
where $F$ is a numerical factor describing the non-linearity of the
thermal conductance between the substrate temperature and the bath
temperature (typically 0.5 for these detectors).

The saturation power of the detectors is dependent on the thermal
conductance as:
  \begin{equation}
		\label{eqn:satpow}
    P_\mathrm{sat}=G_0 T_0 \frac{(T_c/T_0)^{\beta+1}-1}{\beta+1}
  \end{equation}
where the exponent $\beta$ is roughly 2.5
for these detectors.  For the \keckarray\, the
loading from the optics and the sky was modeled to be $\sim$22~\krjtxt.
The optical efficiency is used to convert the loading temperature to a
power deposited on the detector.  For the median optical efficiency
described in \S\ref{sec:oe}, this corresponds to
$P_\mathrm{opt}=$3.1~pW of loading under normal observing conditions.
Assuming a safety factor of 2, the optimal $G_c$ is then 67~pW/K.

The thermal conductance $G_c$ was measured using detector load curves
with the substrate held at different temperatures.  This
method used ``dark'' detectors that
were purposefully disconnected from their antennas to
avoid the optical loading effects.  The detectors used in
\biceptwo\ had higher $G_c$, with two tiles centered at 100~pW/K and
two centered at 140~pW/K.  The tiles fabricated in later runs for the
\keckarray\ had lower thermal conductances, with a median $G_c$ of 90~pW/K.
Several tiles had a much lower $G_c$ of
30--50~pW/K, expected to give lower phonon noise but a smaller margin of
safety against saturation.

Finally, the margin of safety can be verified by measuring the
electrical power $P_\mathrm{J}$ required to keep the detector in transition
during standard observation.  The standard observing schedule includes load
curves (bolometer current-voltage measurements) taken once per hour.
These have been used retrospectively to assess the safety margin under actual
atmospheric conditions.  With the telescope pointed at
55$\deg$ in elevation, the detectors were found to have a median margin of safety
of 6.8~pW, corresponding to a safety factor of 3.2.

\subsection{Beams}
\label{sec:beams}

The beam shapes were measured {\it in situ} at the South Pole by
scanning on a large thermal noise source mounted $\sim 200$~m away,
in the optical far field.
All receivers in the \keckarray\ have a beam width of 0.22~degrees, with
very low levels of ellipticity.  As in \biceptwo\, the dominant differential
beam imperfection for the \keckarray\ is differential
pointing.  The beam mapping campaign, extracted beam parameters, and
residual beam features are described in detail in the Beams Paper.
In this paper we use the high-fidelity per-detector beam maps as a
convolution kernel for simulations to place a limit on the false
$B$-mode signal from beam imperfections.  The simulations and results are
described in \S\ref{sec:systematics}.

\section{Observations and data set}
\label{sec:obs}

\subsection{Observations}
\label{sec:obsstrat}

The observation strategy of the \keckarray\ in the 2012 and 2013
seasons was very similar to that used by \biceptwo, as described
in the \biceptwo\ Results and Instrument Papers.
The same field as \biceptwo\ (and \bicepone) was
observed---a region centered at RA 0h, Dec.\ $-57.5\deg$.
As viewed from the South Pole, the observing field remains at constant
elevation and rotates in azimuth once per day.
For fifty minutes periods
the telescope scanned in azimuth at a fixed elevation, forming
a ``scanset'' with 102 half-scans.
Between scansets,
the azimuth was updated by approximately 12.5$\deg$ to account for the sky rotation,
and stepped in elevation by $0.25\deg$.
Before and after each
scanset, an elevation nod was performed to calibrate the relative gain of the detectors.
The scan rate in azimuth was $2.8\deg/s$.

A group of ten scansets over successive azimuth ranges (and stepping
in elevation) is called a ``phase''.
Table~\ref{tab:phase} shows the phases for the \keckarray.
The elevation ranges were switched between phases after each full cycle of schedules.
Since the briefest sub-kelvin hold time among the set of five helium sorption refrigerators
was $\sim$48 hours, the standard observing schedule consisted of
four CMB phases and one galaxy phase between fridge cycles.

\begin{table}[t]
\caption{The \keckarray\ observation phases}
\label{tab:phase}
\begin{center}
\begin{tabular}{lccll} 
\hline
\hline
\rule[-1ex]{0pt}{3.5ex} Phase & LST Time & Field & Elevation [deg] & Azimuth [deg]\\
\hline
A & Day 0 23:00 & Cryo service & & \\
B & Day 1 05:30 & CMB          & 55.00--57.25 & 120--300 \\
C & Day 1 14:30 & CMB          & 57.50--59.75 & -10--170\\
D & Day 1 23:00 & Galaxy       & 55.00--56.50 & 130--270\\
E & Day 2 05:30 & CMB          & 57.50--59.75 & 120--300\\
F & Day 2 14:30 & CMB          & 55.00--57.25 & -10--170 \\
  \hline
\end{tabular}
\end{center}
\end{table}

As for \biceptwo, the \keckarray\ mount allows for rotation
of the whole apparatus around the line of sight---referred
to as ``deck rotation''.
This rotation was performed between each two day schedule.
For \biceptwo\ and the \keckarray\ in 2012, four deck angles were
used: 68, 113, 248, and 293.
These four angles provide coverage in $Q$ and $U$ and allow for
cancellation of systematic effects whose sign reverses under
180$\deg$ rotation.
In 2013, the \keckarray\ started observing at eight
deck angles: 23, 68, 113, 158, 203, 248, 293, and 338$\deg$.
This allows for a more complete cancellation of beam systematics---see
the Systematics Paper.

\subsection{Data selection}
\label{sec:cuts}

As described in the \biceptwo\ Results and Instrument Papers,
data cuts are applied in three distinct stages.
A few cuts remove half-scans from the scansets, while a larger number cut
entire scansets from the final map.
The final cut stage is the channel selection cut
which is applied during the final coaddition stage.
These three stages provide the necessary flexibility and granularity.
The cuts are summarized in Table~\ref{tab:cuts} for the
2012--2013 data set (Cf.\ Table~7 of the \biceptwo\
Instrument Paper).

\begin{table*}[t]
\caption{Data selection cuts for the \keckarray\ for 2012--2013}
\label{tab:cuts}
\begin{center}
\begin{tabular}{lccc} 
\hline
\hline
\rule[-1ex]{0pt}{3.5ex} Cut & Total time [10$^6$ s] & Integration [$10^9$ det$\cdot$s] & Fraction cut [\%]\\
\hline
  Before cuts & 18.3 & 30.4 &  \\
  Channel cuts & 18.3 & 27.5 & 9.5 \\
  Synchronization & 18.1 & 27.2 & 1.2 \\
  Deglitching & 17.8 & 23.6 & 11.8 \\
  Passing channels 1 & 17.6 & 23.3 & 0.58 \\
  Elnod calibration & 17.1 & 19.9 & 11.2 \\
  TES fractional resistance & 17.1 & 19.6 & 0.88 \\
  Time stream skewness & 17.1 & 17.8 & 6.2 \\
  Time stream variance & 17.0 & 17.4 & 1.2 \\
  Noise stationarity & 16.9 & 17.0 & 1.4 \\
  FPU Temperature & 16.7 & 16.7 & 0.84 \\
  Telescope pointing & 15.7 & 15.8 & 3.0 \\
  Passing data & 15.4 & 15.6 & 0.55 \\
\hline

\end{tabular}
\end{center}
\end{table*}

\subsection{2012--2013 data set and sensitivity}
\label{sec:sens}

\begin{figure*}
\resizebox{\textwidth}{!}{\includegraphics{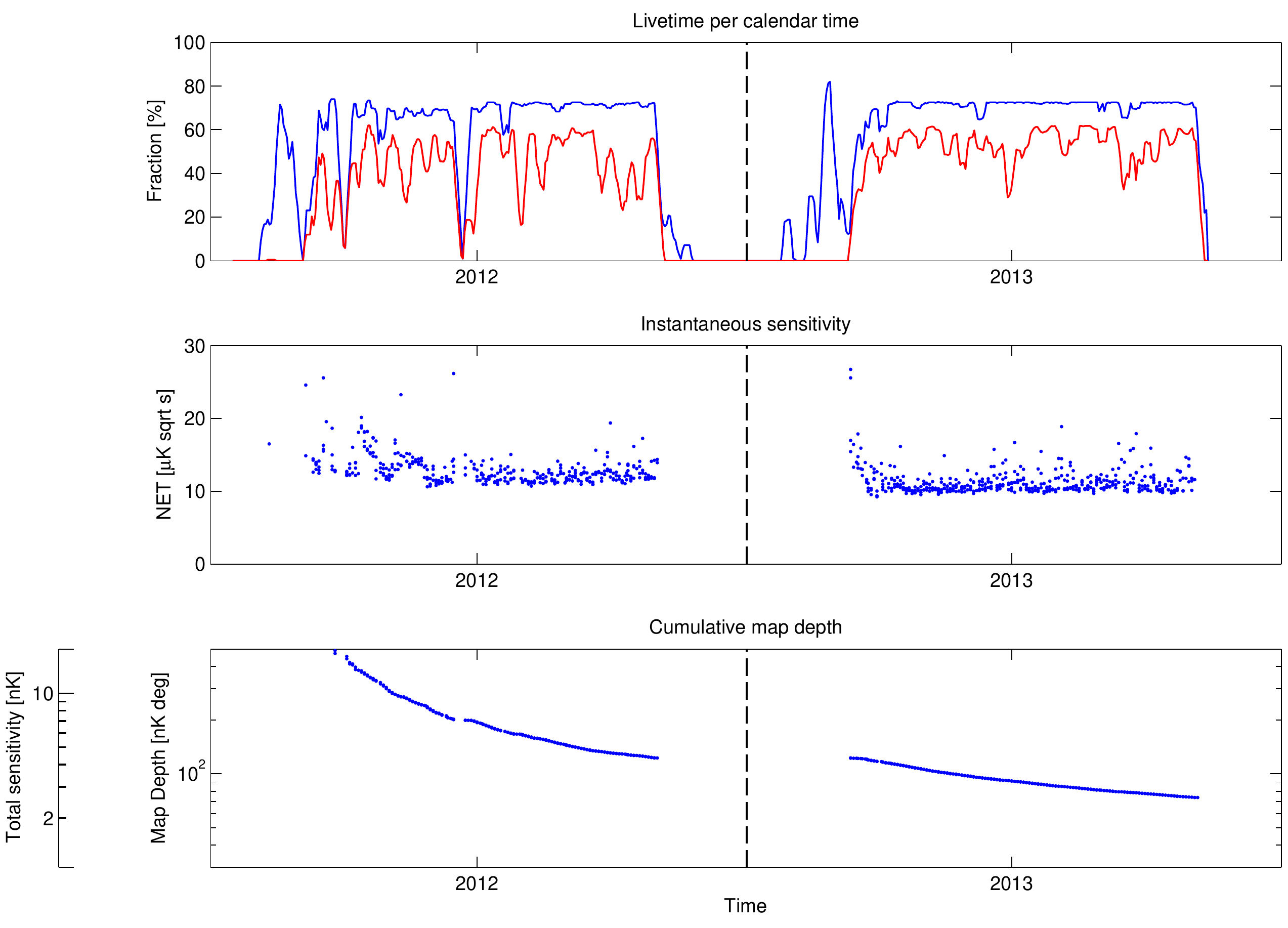}}
\caption{The \keckarray\ 2012--2013 150~GHz data set.  The top panel
shows the fraction of calendar time the telescopes were observing.  The
red line represents the fraction of time preserved after all data selection
cuts are implemented. The middle panel shows the instantaneous sensitivity
of the full \keckarray. The bottom panel shows the cumulative map depth
as calculated over each phase (10~hr of data).}
\label{fig:livetime}
\end{figure*}

The telescopes continuously took data through the South Pole winter.
Each of 2012 and 2013 produced
nearly 4500 scansets of data for a total of $18\times10^6$~s of data.
After the data selection cuts, an average of 50\% of the data remain,
for a total of 15.6$\times 10^9$ dets$\cdot$s.  This is
shown graphically in the top panel of Figure~\ref{fig:livetime}
(Cf.\ Figure~23 of the \biceptwo\ Instrument Paper).

The instantaneous sensitivity of the \keckarray\ instruments
is measured using two methods: by taking the average of the
time stream noise spectra between 0.1--1~Hz and by measuring the standard
deviation of the noise-only maps weighted by the square root of
integration time~\citep{kernasovskiy12}.  Both methods yielded a
noise-equivalent-temperature (NET) of 11.5~\ukcmbrts for 2012
and 9.5~\ukcmbrts for 2013.  The middle panel of Figure~\ref{fig:livetime}
shows the instantaneous sensitivity calculated with the time stream
based method for the 2012--2013 seasons.
Using the same method as described in the \biceptwo\ Results and
Instrument Papers the map depth for the \keckarray\ 2012--2013 150~GHz
data is 74~nK in nominal square-degree pixels (4.4~$\mu$K$\,$arcmin)
over an effective area of 390 square degrees for a total
sensitivity of 2.6~nK.  An equivalent way of expressing the sensitivity
of the data set is the survey weight $W=1/s^2=$150,000~$\mu$K$^{-2}$,
where $s$ is the total sensitivity.
This expression is useful because it scales linearly with
integration time, number of detectors, and statistical sensitivity
to $r$.

\section{Low level data reduction, map making and simulations}

\subsection{Analysis pipeline}
\label{sec:pipeline}

The \keckarray\ and \biceptwo\ data analysis uses the
same code and proceeds in parallel, enabling cross
checks between these independent data sets.
The process used here is exactly the same
as described in the \biceptwo\ Results Paper---a summary follows.

\subsection{Low level reduction}
\label{sec:lowreduc}

As detailed in the \biceptwo\ Results Paper, the first step in the low
level reduction is to deconvolve the temporal response of the instrument.
The TES detectors have a fast response of $\sim$1~ms that can be
ignored.  Both the MCE and the \gcp\ apply low pass filters
to the data which must be accounted for in the
deconvolution.

At this stage, relative calibration is accomplished by
dividing the time streams by the individual detector
gains derived from elevation nods.  The data are then multiplied by
the median gain across the array in order to remove dependence
on atmospheric variation.

\subsection{Pairmaps}

The sum and difference of each detector pair is taken.
Each half-scan is subjected to third order polynomial
filtering to remove atmospheric variation.
In addition, the mean of the half-scans over the
scanset is subtracted to remove any
scan-synchronous contamination.
The pointing of each detector pair is reconstructed
from the telescope pointing model and per-pair
offset angles refined by regressing per-channel maps
against a \wmap5\ template.
At this point the time stream data of each pair
is binned into a rectangular grid of pixels forming
per scanset ``pairmaps''.
We also sample and bin the \planck\ 143~GHz temperature map and
its derivatives to use in the deprojection of beam
systematics.
For further details see the \biceptwo\ Results Paper
and the Systematics Paper.

\subsection{Full maps}

Finally, the pairmaps are coadded into final full maps, and also
into various pairs of jackknife splits.
The deprojection templates are fit and removed during this process.
Absolute calibration of the maps is performed by comparing the
power spectrum of the temperature map with the \planck\ 143~GHz map
as described in the Instrument Paper.

\begin{figure*}
\resizebox{\textwidth}{!}{\includegraphics{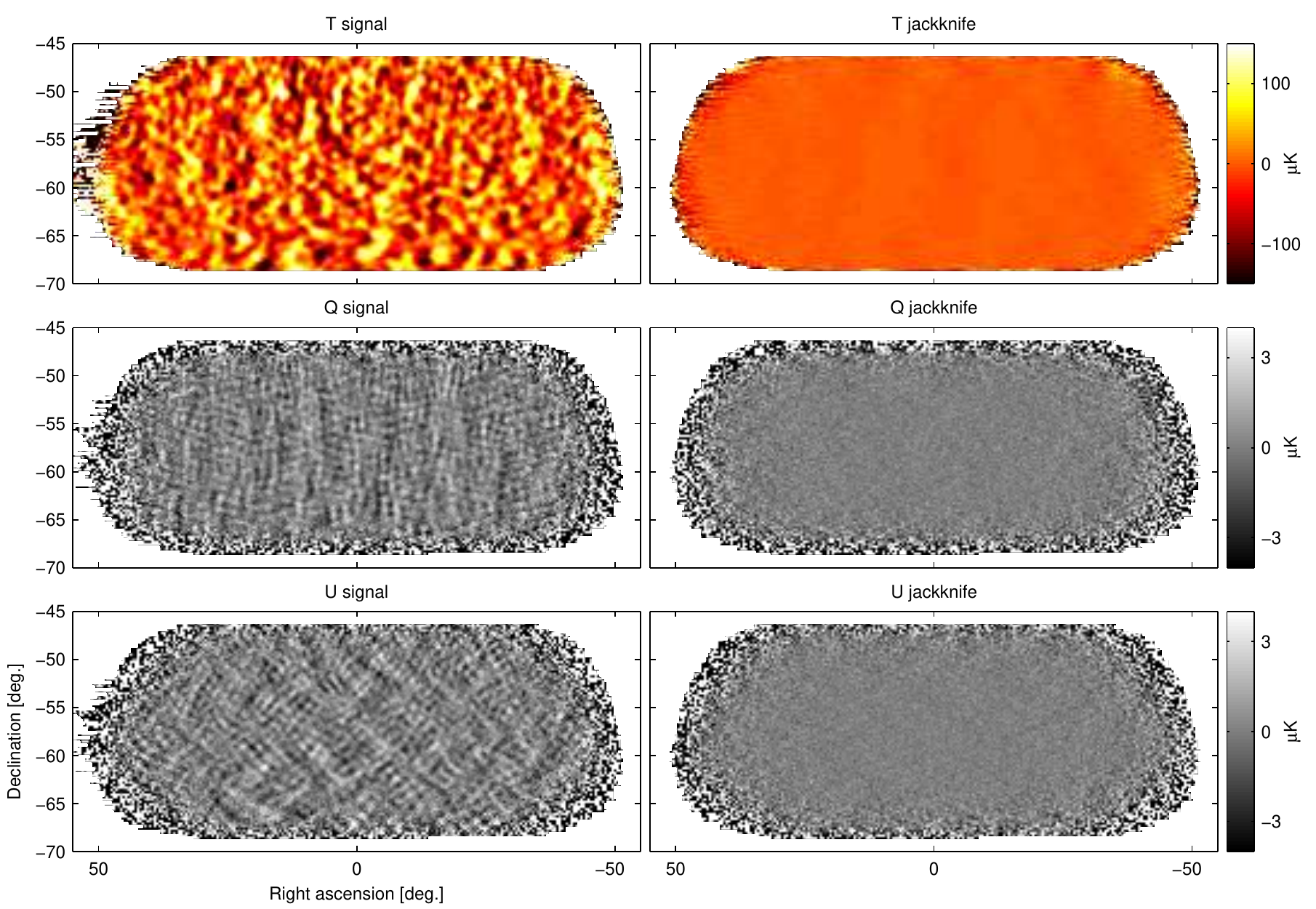}}
\caption{\keckarray\ $T$, $Q$, $U$ maps.
The left column shows the basic signal maps with $0.25\deg$ pixelization
as output by the reduction pipeline.
The right column shows difference (jackknife) maps made with the
first halves of the 2012 and 2013 seasons and the second halves.
No additional filtering other than that imposed by the instrument beam
(FWHM $0.5\deg$) has been done.
Note that the structure seen in the $Q$ and $U$ signal maps is as expected
for an \emode\ dominated sky.}
\label{fig:tqu_maps}
\end{figure*}

Figure~\ref{fig:tqu_maps} shows the resulting temperature and Stokes
$Q$ and $U$ polarization maps for the \keckarray\ 150~GHz data from the
2012--2013 observing seasons.
The left column shows the final maps and
the right hand column shows a difference (jackknife) map
which is consistent with noise alone.
The vertical/horizontal stripes in the $Q$ maps and the diagonal stripes
in the $U$ maps are characteristic of the \emode\ polarization signal,
which dominates the maps.

\subsection{Simulations}
\label{sec:sims}

We create signal and noise simulations exactly as described in
the \biceptwo\ Results Paper.
We generate realizations of noise by randomly flipping the signs
of the pairmaps when co-adding to full maps.
Several kinds of signal simulations are made by resampling
input maps from the \synfast\ program (part of the \healpix\footnote{
\url{http://healpix.sourceforge.net/}} package~\citep{healpix}).
The simulated data are then binned into pairmaps and
combined to full maps exactly in parallel with the
treatment of the real data.

\section{Results}

\subsection{Power spectra}
\label{sec:powspec}

The maps are converted into angular power spectra
exactly as described in the \biceptwo\ Results Paper.
The matrix based purification of the $Q$ and $U$ maps
is performed prior to inversion to form {\bmode}s
to avoid $E$ to $B$ mixing due to the sky-cut and filtering.
A variant of the \master\ procedure~\citep{hivon02} is
used to noise debias the auto spectra and correct for
the effects of the time stream filtering.

\begin{figure*}
\resizebox{\textwidth}{!}{\includegraphics{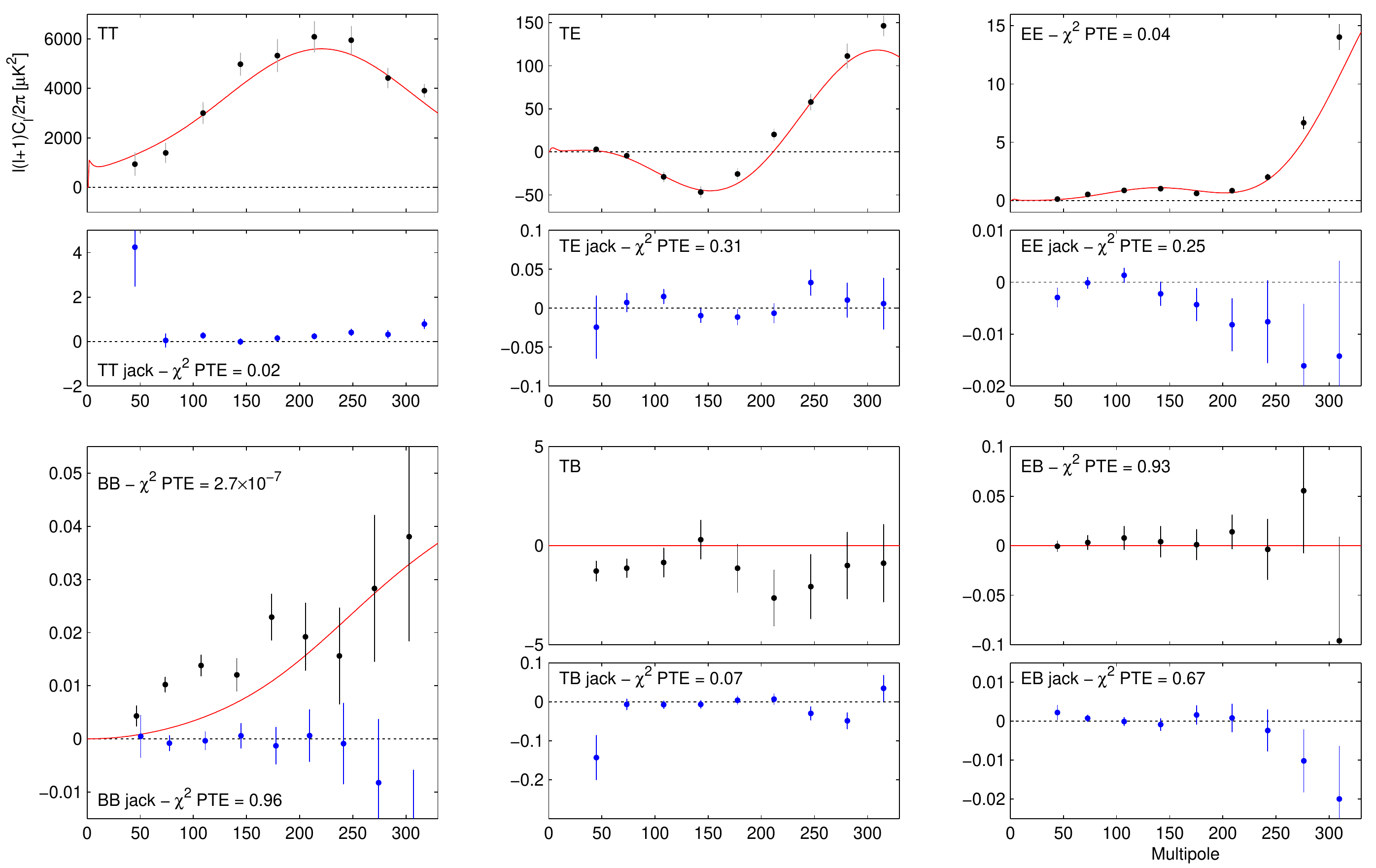}}
\caption{\keckarray\ power spectrum results for signal (black points)
and early/late season jackknife (blue points).
The solid red curves show the lensed-\lcdm\ theory expectations.
The error bars are the standard deviations of the
lensed-\lcdm+noise simulations and hence contain no sample variance
on any additional signal component.
The probability to exceed (PTE) the observed value of a simple
$\chi^2$ statistic for the 9 band powers is given (as evaluated against the simulations).
Note that the band powers of the auto spectra of the simulations are
approximately $\chi^2$ distributed, with the lowest $\ell$-bin only containing
$\sim$9 effective degrees of freedom.  This increases the probability of
an outlier point in comparison to a Gaussian distribution.  The observed
distribution for the cross spectra are more symmetric than the $\chi^2$ distributions
but have similarly increased tails.
This is fully reflected in the quoted PTE value.
Also note the very different $y$-axis scales for the jackknife spectra
(other than $BB$).
See the text for additional discussion of the $BB$ spectrum.
(Note that the calibration procedure uses $EB$ to set the overall polarization angle
so $TB$ and $EB$ as plotted above cannot be used to measure astrophysical
polarization rotation.)}
\label{fig:powspecres}
\end{figure*}

The resulting power spectra for the \keckarray\ using the 150~GHz data from the 2012--2013
observing seasons is shown in Figure~\ref{fig:powspecres},
along with a temporal jackknife.
The \bmode\ power spectrum is inconsistent with \lcdm\
cosmology without foregrounds at $5.0\sigma$ (over the first 5 band powers).
Although the noise is lower than for \biceptwo, the first two band power
values are also lower so the significance is somewhat smaller.

An overall rotation is applied to the maps to minimize
the high-$\ell$ $TB$ and $EB$ spectra~\citep{kaufman14}.
For the \keckarray\ 2012+2013 data this adjustment is $\approx -0.5\deg$.
This rotation makes no practical difference to the \bmode\ power spectrum.

\subsection{$E$ and $B$ maps}

\begin{figure*}
\resizebox{\textwidth}{!}{\includegraphics{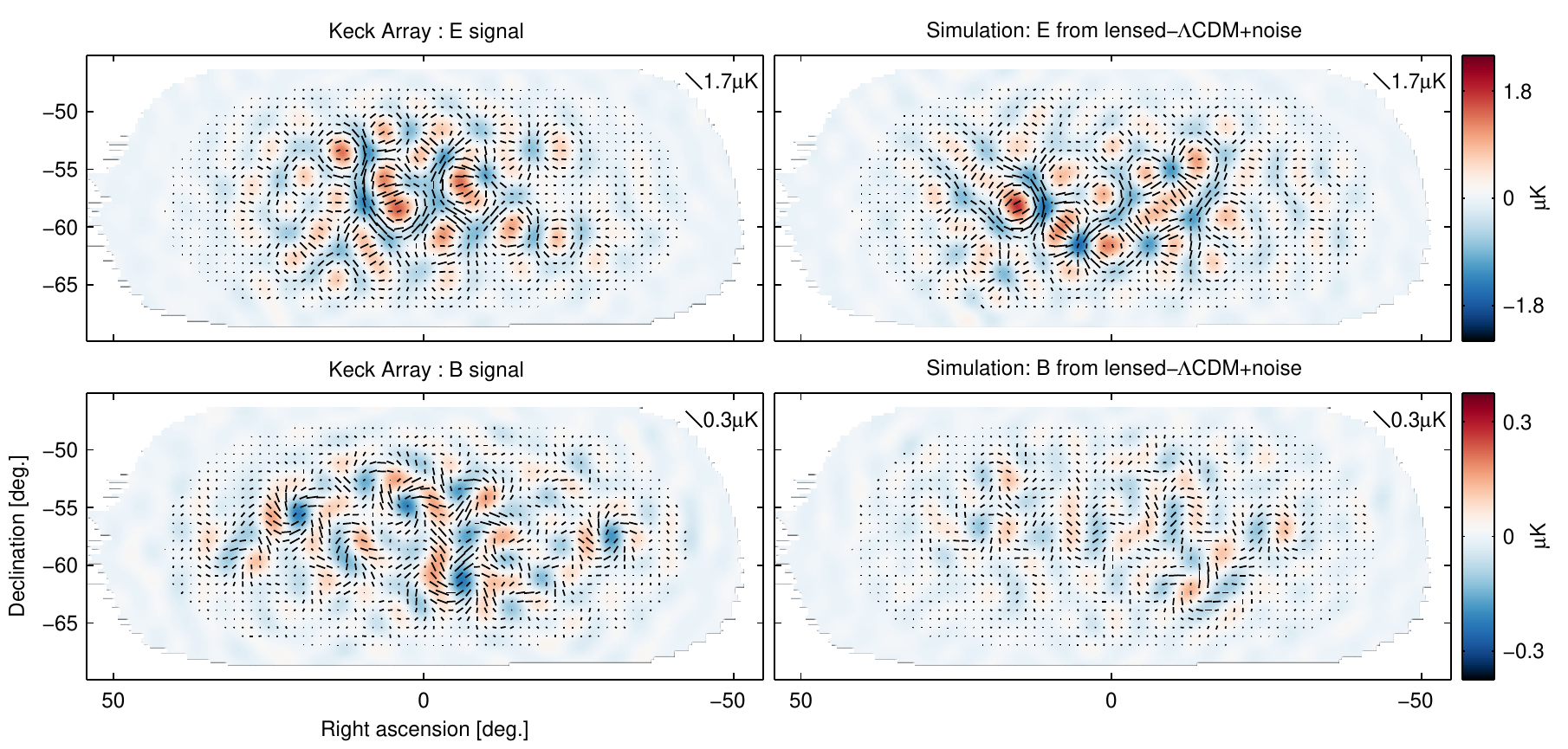}}
\caption{{\it Left:} \keckarray\ apodized \emode\ and \bmode\ maps filtered to $50<\ell<120$.
{\it Right:} The equivalent maps for the first of the lensed-\lcdm+noise simulations.
The color scale displays the \emode\ scalar and \bmode\ pseudoscalar patterns
while the lines display the equivalent magnitude and orientation of linear polarization.
Note that the \emode\ and \bmode\ maps use different
color/length scales.}
\label{fig:ebmaps}
\end{figure*}

\emode\ and \bmode\ maps can be made by performing an inverse
Fourier transform as shown in Figure~\ref{fig:ebmaps}.  The maps
created are inherently apodized, as the \emode\ and \bmode\
components are generated from apodized Q and U maps.  These
are compared to a lensed-\lcdm+noise simulation.

\subsection{Internal consistency tests}
\label{sec:jacks}

The \keckarray\ data was split in 16 different ways to test for internal
consistency.
The motivations behind these splits are described in the
\biceptwo\ Results Paper and the Systematics Paper.
If a contaminating signal exists in only one half of the data split,
then it should show up with as much significance in the jackknife
as in the signal map.
However, some jackknives are more sensitive to certain
systematics than the signal map because of inherent
cancellation effects which operate in the full map.
Each of the jackknife categories is summarized below.

The first set of jackknives probes for systematics which differ
between different subsets of channels.
This includes division in the multiplexing
system, as well as divisions in the focal plane layout: tile,
focal plane inner/outer, tile inner/outer, mux row, mux column.
As is documented in the Beams Paper, there are systematics that
are highly dependent on the position of the detector in the focal plane.
For instance, the ellipticity of the beams is greater in the detectors
near the outside of the focal plane than the inside.

The next set of jackknives is temporal. This includes both the
longest time scale of 2012 data versus 2013 data, and the shortest
timescale of left going scans versus right going scans.
Owing to the changes between the 2012 and 2013 observing seasons,
an early/late season jackknife acts as an alternative temporal split.
The first is sensitive to the effects of different observing schedules and
detectors changed between seasons, while the second is sensitive
to detector time constants.

Another set of jackknives is based on external contamination.  This
includes the azimuth jack, which divides the data based on the direction
the telescope is pointed with respect to the ground (see Table~\ref{tab:phase}).
One half of this jackknife is data taken in the direction
of the main South Pole Station and associated operations, while
the other half points into the desolate Antarctic plateau.

A set of jackknives that particularly amplifies the differential beam properties is the
deck rotation jackknives.
As is described in the \biceptwo\ Systematics Paper, a 180$^\circ$
deck rotation cancels out differential pointing.
The deck jackknife, which differences the 180$^\circ$ rotations,
amplifies the leakage by an order of magnitude in comparison
to that present in the fully coadded data.
The \keckarray\ also started taking data at 90$^\circ$ compliment
deck rotations in the 2013 observing season, and this jackknife
is sensitive to differential gain or differential beam width leakage.
The alternative deck jackknife is defined to be the difference of the 90$^\circ$
rotations for 2013.
In this special case, the statistics for the 2012 and 2013 data
are separate.

Maps are made from each half of the data split and then
differenced.  The differenced maps are divided by a factor
of two in order to keep the noise amplitudes equivalent to the signal map.
The consistency with lensed-\lcdm+noise
simulations is calculated with a simple $\chi^2$ statistic:
\begin{equation}
\chi^2 = \left( \mathbf{d} - \langle\mathbf{m}\rangle\right)^\mathrm{T} \mathbf{D}^{-1}
\left(\mathbf{d} - \langle\mathbf{m}\rangle\right)
\label{eqn:chisq}
\end{equation}
where $\mathbf{d}$ is the vector of observed band power values,
$\langle\mathbf{m}\rangle$ is
the mean of the lensed-\lcdm+noise simulations (except where
alternative signal models are considered), and $\mathbf{D}$
is the band power covariance matrix as evaluated from those
simulations.

A $\chi$ statistic is also considered to probe for sets
of band powers which are systematically above or
below the expectation.
This is defined as:
\begin{equation}
\chi = \sum_i \frac{d_i - \langle m_i \rangle}{\sigma_{m_i}}
\end{equation}
where the $d_i$ are the observed band power values and $\langle m_i \rangle$
and $\sigma_{m_i}$ are the mean and standard deviation of the
lensed-\lcdm+noise simulations.

\begin{figure}
\begin{center}
\resizebox{\columnwidth}{!}{\includegraphics{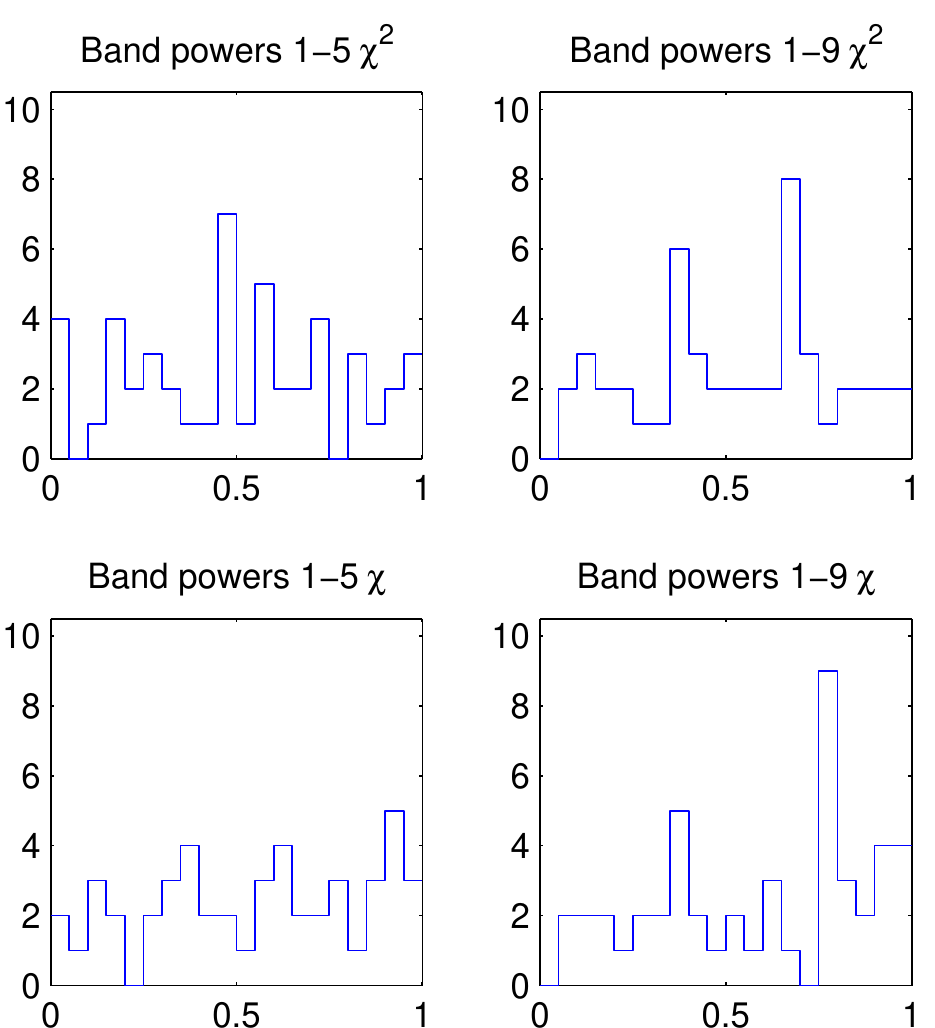}}
\end{center}
\caption{Distributions of the jackknife $\chi^2$ and $\chi$ PTE
values over the tests and spectra given in Table~\ref{tab:ptes}.}
\label{fig:ptedist}
\end{figure}

For each of these statistics, we calculate the probability
to exceed (PTE) the observed value by comparing to the values
obtained in the 500 lensed-\lcdm+noise simulations.
Since the distribution of the band powers of the auto spectra
is approximately $\chi^2$ distribution, there is some
non-Gaussianity to the statistics.  In particular, the lowest band
power only has 9 effective modes which will increase the tails
of the distribution.  However, by calculating the PTE against the
simulations, any non-Gaussianity is fully reflected in the PTE value.
The PTE for the $\chi$ and $\chi^2$ using band powers 1--5 and 1--9
is given in Table~\ref{tab:ptes}.
Note that these statistics are correlated (especially along
each row of the table).
The distribution of the PTE values is shown in Figure~\ref{fig:ptedist}.

\begingroup
\squeezetable
\begin{table}[ht]
\caption{Jackknife PTE values from $\chi^2$ and $\chi$ (sum of deviation) tests \label{tab:ptes}}
\begin{ruledtabular}
\begin{tabular}{l c c c c }
Jackknife & Band powers & Band powers & Band powers & Band powers\\
& 1--5 $\chi^2$ & 1--9 $\chi^2$ & 1--5 $\chi$ & 1--9 $\chi$ \\
\hline
\\ 
\multicolumn{5}{l}{Deck jackknife} \\ 
EE & 0.613 & 0.924 & 0.898 & 0.776 \\ 
BB & 0.743 & 0.880 & 0.685 & 0.375 \\ 
EB & 0.820 & 0.986 & 0.309 & 0.429 \\ 
\multicolumn{5}{l}{Scan Dir jackknife} \\ 
EE & 0.561 & 0.415 & 0.788 & 0.898 \\ 
BB & 0.924 & 0.691 & 0.601 & 0.180 \\ 
EB & 0.168 & 0.453 & 0.938 & 0.886 \\ 
\multicolumn{5}{l}{Early/Late Season jackknife} \\ 
EE & 0.287 & 0.255 & 0.896 & 0.998 \\ 
BB & 0.982 & 0.960 & 0.621 & 0.796 \\ 
EB & 0.711 & 0.667 & 0.170 & 0.609 \\ 
\multicolumn{5}{l}{Year Split jackknife} \\ 
EE & 0.343 & 0.641 & 0.918 & 0.956 \\ 
BB & 0.856 & 0.940 & 0.695 & 0.657 \\ 
EB & 0.747 & 0.547 & 0.353 & 0.798 \\ 
\multicolumn{5}{l}{Tile jackknife} \\ 
EE & 0.042 & 0.110 & 0.431 & 0.782 \\ 
BB & 0.573 & 0.715 & 0.118 & 0.499 \\ 
EB & 0.451 & 0.691 & 0.940 & 0.932 \\ 
\multicolumn{5}{l}{Phase jackknife} \\ 
EE & 0.826 & 0.824 & 0.743 & 0.309 \\ 
BB & 0.036 & 0.184 & 0.489 & 0.343 \\ 
EB & 0.026 & 0.058 & 0.980 & 0.914 \\ 
\multicolumn{5}{l}{Mux Col jackknife} \\ 
EE & 0.804 & 0.142 & 0.543 & 0.080 \\ 
BB & 0.471 & 0.760 & 0.291 & 0.295 \\ 
EB & 0.144 & 0.206 & 0.585 & 0.840 \\ 
\multicolumn{5}{l}{Alt Deck jackknife 2012} \\ 
EE & 0.673 & 0.884 & 0.641 & 0.377 \\ 
BB & 0.579 & 0.685 & 0.569 & 0.784 \\ 
EB & 0.152 & 0.112 & 0.389 & 0.609 \\ 
\multicolumn{5}{l}{Alt Deck jackknife 2013} \\ 
EE & 0.489 & 0.583 & 0.269 & 0.507 \\ 
BB & 0.583 & 0.822 & 0.794 & 0.826 \\ 
EB & 0.549 & 0.441 & 0.577 & 0.575 \\ 
\multicolumn{5}{l}{Mux Row jackknife} \\ 
EE & 0.942 & 0.645 & 0.421 & 0.776 \\ 
BB & 0.363 & 0.371 & 0.625 & 0.786 \\ 
EB & 0.214 & 0.345 & 0.030 & 0.112 \\ 
\multicolumn{5}{l}{Tile/Deck jackknife} \\ 
EE & 0.204 & 0.052 & 0.910 & 0.240 \\ 
BB & 0.487 & 0.745 & 0.986 & 0.830 \\ 
EB & 0.491 & 0.351 & 0.146 & 0.603 \\ 
\multicolumn{5}{l}{Focal Plane inner/outer jackknife} \\ 
EE & 0.253 & 0.475 & 0.108 & 0.064 \\ 
BB & 0.637 & 0.587 & 0.074 & 0.158 \\ 
EB & 0.044 & 0.226 & 0.996 & 0.994 \\ 
\multicolumn{5}{l}{Tile top/bottom jackknife} \\ 
EE & 0.573 & 0.397 & 0.924 & 0.910 \\ 
BB & 0.289 & 0.194 & 0.198 & 0.782 \\ 
EB & 0.172 & 0.353 & 0.884 & 0.954 \\ 
\multicolumn{5}{l}{Tile inner/outer jackknife} \\ 
EE & 0.707 & 0.663 & 0.399 & 0.387 \\ 
BB & 0.303 & 0.663 & 0.719 & 0.786 \\ 
EB & 0.958 & 0.655 & 0.315 & 0.102 \\ 
\multicolumn{5}{l}{Moon jackknife} \\ 
EE & 0.192 & 0.433 & 0.395 & 0.385 \\ 
BB & 1.000 & 0.387 & 0.339 & 0.505 \\ 
EB & 0.667 & 0.705 & 0.794 & 0.289 \\ 
\multicolumn{5}{l}{A/B offset best/worst} \\ 
EE & 0.483 & 0.521 & 0.804 & 0.926 \\ 
BB & 0.443 & 0.367 & 0.042 & 0.407 \\ 
EB & 0.497 & 0.677 & 0.489 & 0.397 \\ 

\end{tabular}
\end{ruledtabular}
\end{table}
\endgroup

\section{Systematics}
\label{sec:systematics}

Experimental systematics can create false \bmode\ polarization
and must be shown to be tightly controlled.
The systematics in \biceptwo\ were
fully explored in the Systematics Paper and were shown to be
below the level equivalent to $r$=0.003--0.006.
These limits were derived from forward simulations of the
measured instrumental properties.
If a given property did not have a measured level, appropriate upper
limits were used.

The beam systematics in the \keckarray\ are
expected to be below those of \biceptwo\ because of the larger number
of detectors (increased averaging down of incoherent effects),
and the increased number of receiver orientations
(both instantaneously due to the ``clocking'' of the five receivers
at $72^\circ$ increments around the boresight, and, in 2013, the
increased number of deck angles of observation).

\begin{figure}
\begin{center}
\resizebox{\columnwidth}{!}{\includegraphics{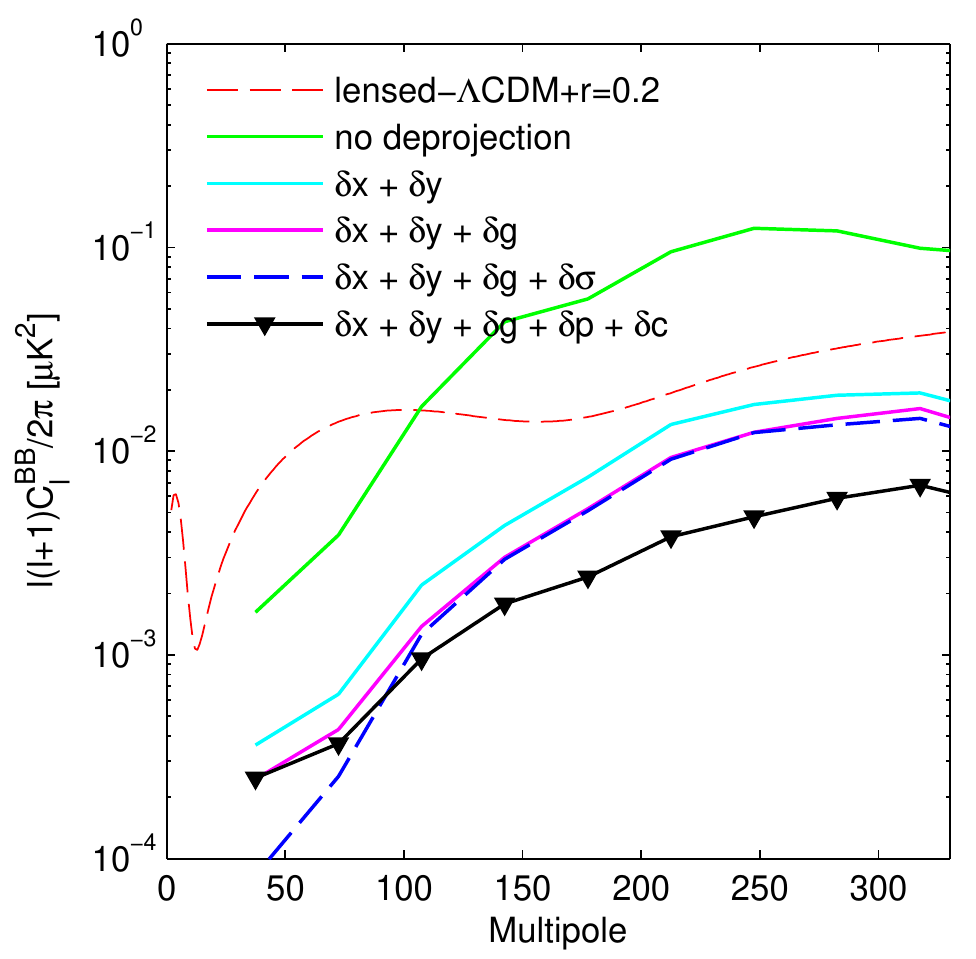}}
\end{center}
\caption{$BB$ spectra from $T$-only input simulations using
the measured per channel beam shapes
compared to the lensed-\lcdm+$r=0.2$ spectrum.
From top to bottom the curves are (i) no deprojection,
(ii) deprojection of differential pointing only ($\delta x + \delta y$),
(iii) deprojection of differential pointing and differential gain of the
detector pairs ($\delta x + \delta y + \delta g$),
(iv) adding deprojection of differential beam width ($\delta x + \delta y +
\delta g + \delta\sigma$), and
(v) differential pointing, differential gain, and differential
ellipticity ($\delta x + \delta y + \delta g + \delta p + \delta c$).
This last curve represents an upper limit only to the residual contamination.}
\label{fig:beamsim}
\end{figure}

As described in the \biceptwo\ Results and Systematics Papers
we produce simulated time streams by convolving
an input temperature map with high precision per
channel measurements of the actual beam shapes.
We then pass these simulated time streams through the mapping
process, including all filtering and deprojection, to assess the residual
contamination due to beam non-ideality.
The results are shown in Figure~\ref{fig:beamsim}.
The beam maps for the \keckarray\ do not provide as uniform and
redundant coverage of all detectors as those for \biceptwo\, and
additional analysis is required to construct composite beam maps
that have consistently high signal-to-noise and are free of artifacts from the
beam mapping measurement.
For the purposes of the current paper we use the preliminary beam map results
to set an upper limit
on the residual contamination, as indicated by the
down arrows in the figure.

Other forms of systematic contamination were considered, such as
electromagnetic interference (EMI) contamination,
magnetic pickup, thermal pickup, and detector
pointing.  Each of these systematics was quantified to be below
the \biceptwo\ level when averaged over the entire array, and thus
safely ignorable.

\section{Consistency with \biceptwo}
\label{sec:b2consis}

\begin{figure}
\begin{center}
\resizebox{\columnwidth}{!}{\includegraphics{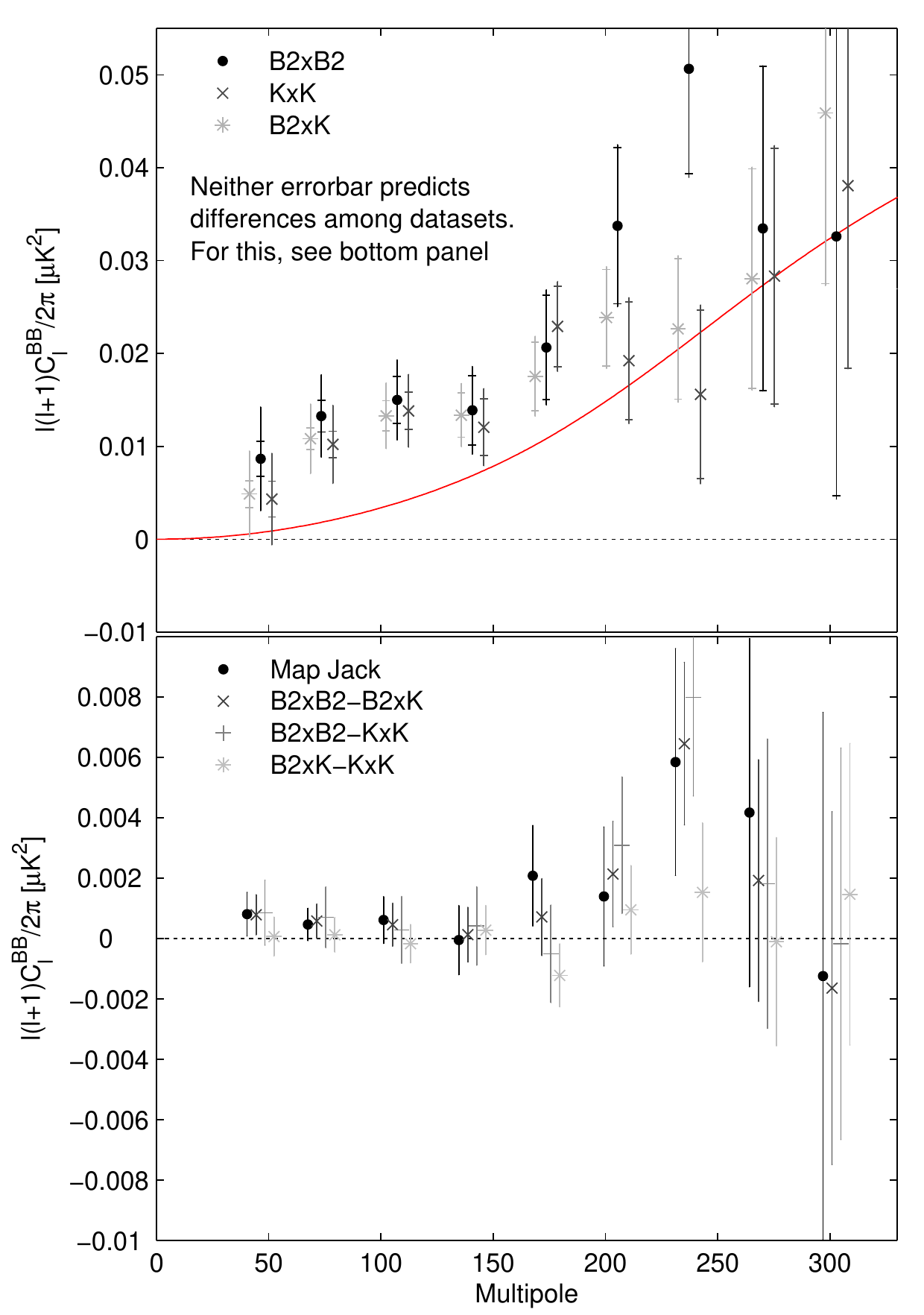}}
\end{center}
\caption{{\it Upper:} The \keckarray\ $BB$ auto spectrum,
the \biceptwo\ auto spectrum, and the cross spectrum taken between the two.
The inner error bars are the standard deviation of the lensed-\lcdm+noise simulations,
while the outer error bars also contain excess power at low-$\ell$.
(For clarity the \keckarray\ and cross spectrum points
are offset horizontally.)
{\it Lower:}
Four compatibility tests between the {\bmode}s measured by
\biceptwo\ and \keckarray.
The ``map jack'' takes the difference of the $Q$ and $U$ maps,
divides by a factor of two,
and calculates the $BB$ spectrum.
The other three sets of points are the differences of the spectra shown in
the upper panel divided by a factor of four.
In each case the error bars are the standard deviation of the
pairwise differences of signal+noise simulations
which share common input skies.  Comparison of any one of these
sets of points with null is an appropriate test
of the compatibility of the experiments---see text for details.
All tests show good consistency between \biceptwo\ and \keckarray,
particularly in the lowest five bandpowers.}
\label{fig:specjack}
\end{figure}

The resulting \bmode\ power spectra of \biceptwo\ and the \keckarray\ can be
compared to assess the compatibility of the two sets of results.
Although there is much that the two experiments
share in terms of hardware, design, and
location, there are also potential systematics that are different:
the bulk refrigeration system, the ground shield, and the time at which the
observations occurred.
Comparing the results is a powerful additional systematics check.
The power spectra for both the \keckarray, \biceptwo\, and the cross
between the two are shown in the upper panel of Figure~\ref{fig:specjack}.

A rigorous comparison can be done in two ways: directly comparing the maps and
comparing the auto and cross power spectra.
The latter can be a more powerful comparison if the maps have
different noise levels---since \biceptwo\ and the \keckarray\ have
comparable noise levels, all four methods (map and the three
combinations of auto and cross spectra) have approximately equal statistical power.

To test the compatibility of the resulting band powers
with null we compare them to the differences
of signal+noise simulations which share common input skies.
In such tests it is necessary that the simulations contain power
roughly equal to the real sky as the cross terms between signal
and noise increase the fluctuation of the differences even for perfectly
common sky coverage.  For example,
the (un-debiased) auto spectrum of a map $M$ composed of a signal $S$ and
noise $N$ can be written
\begin{equation}
M \!\! \times \!\! M = (S+N) \! \times \! (S+N)=S \!\! \times \!\! S + 2(S \!\! \times \!\! N) +N \!\! \times \!\! N.
\end{equation}
The difference of such auto spectra between experiments with common sky coverage
is then
\begin{equation}
M_1 \!\! \times \!\! M_1 - M_2 \!\! \times \!\! M_2 =
2(S \!\! \times \!\! N_1 - S \!\! \times \!\! N_2)
+ N_1 \!\! \times \!\! N_1
- N_2 \!\! \times \!\! N_2
\end{equation}
where $M_1$ and $M_2$ refer to the first and second experiment.
The signal auto spectrum $S \!\! \times \!\! S$ cancels out.
However, the cross terms between the signal and noise
$2(S \!\! \times \!\! N_1 - S \!\! \times \!\! N_2)$
do not cancel, and they increase the fluctuations between the two experiments
over the noise-only case in proportion to the common signal.
To account for this extra variance, we use signal simulations
with additional power that matches the amplitude of the
observed signal in excess of \lcdm\ in band powers 1--5.
(The origin of the extra signal over \lcdm\ is not important here---only
its approximate amplitude.)
The results are shown in Figure~\ref{fig:specjack}.

We then proceed to calculate the PTE of the
$\chi$ and $\chi^2$ statistics versus the simulated distributions
using the same spectra and band power ranges
as in~\S\ref{sec:jacks}, and give the results in Table~\ref{tab:specptes}.
In both the figure and the table we note the effect of the two
band powers at $\ell \approx 220$ that are high with
respect to lensed-\lcdm\ in B2xB2 (as noted in the \biceptwo\ Results Paper)
but not in KxK and B2xK---as expected these also show up in
the map difference.
Again note that the PTE values are correlated (both
along and between rows of the table) so overinterpretation
should be avoided.
Our conclusion is that the \biceptwo\ and \keckarray\ data
are consistent---especially in the lowest five band powers
where an IGW contribution would be strongest.

\begingroup
\squeezetable
\begin{table}
\caption{\biceptwo/\keckarray\ compatibility test PTE values from $\chi^2$ and $\chi$
(sum of deviation) tests \label{tab:specptes}}
\begin{ruledtabular}
\begin{tabular}{l c c c c}
Jackknife & Band powers & Band powers & Band powers & Band powers \\
& 1--5 $\chi^2$ & 1--9 $\chi^2$ & 1--5 $\chi$ & 1--9 $\chi$ \\
\hline
\\ 
\multicolumn{5}{l}{Map jackknife} \\ 
EE & 0.034 & 0.048 & 0.106 & 0.028 \\ 
BB & 0.561 & 0.695 & 0.054 & 0.018 \\ 
EB & 0.741 & 0.754 & 0.405 & 0.651 \\ 
\multicolumn{5}{l}{Spectral jackknife B2-cross} \\ 
EE & 0.112 & 0.092 & 0.068 & 0.078 \\ 
BB & 0.687 & 0.387 & 0.052 & 0.008 \\ 
EB & 0.555 & 0.224 & 0.212 & 0.234 \\ 
\multicolumn{5}{l}{Spectral jackknife B2-Keck} \\ 
EE & 0.138 & 0.128 & 0.066 & 0.126 \\ 
BB & 0.920 & 0.485 & 0.200 & 0.022 \\ 
EB & 0.511 & 0.214 & 0.210 & 0.200 \\ 
\multicolumn{5}{l}{Spectral jackknife cross-Keck} \\ 
EE & 0.176 & 0.204 & 0.074 & 0.202 \\ 
BB & 0.880 & 0.966 & 0.643 & 0.435 \\ 
EB & 0.361 & 0.437 & 0.443 & 0.188 \\ 

\end{tabular}
\end{ruledtabular}
\end{table}
\endgroup

\section{Combination with \biceptwo}
\label{sec:comb}

Having shown that the \keckarray\ results are consistent
with \biceptwo\, we now proceed to combine the maps
by adding the accumulation quantities (equivalent to
a noise weighted combination of the maps).
This results in $Q$ and $U$ maps which have a depth of 57~nK$\,$deg
(3.4~~$\mu$K$\,$arcmin) over an effective area of 400 square degrees.
Following \S\ref{sec:sens}, the map depth and effective area are
combined for a total sensitivity of 2.0~nK and
a total survey weight of 250,000~$\mu$K$^{-2}$.

The observation regions and strategies are sufficiently
similar that it is found empirically using simulations
that the purification matrix of either experiment delivers
adequate \bmode\ purity when applied to the combined map
(with contamination equivalent to $r<10^{-3}$).

\begin{figure}
\begin{center}
\resizebox{\columnwidth}{!}{\includegraphics{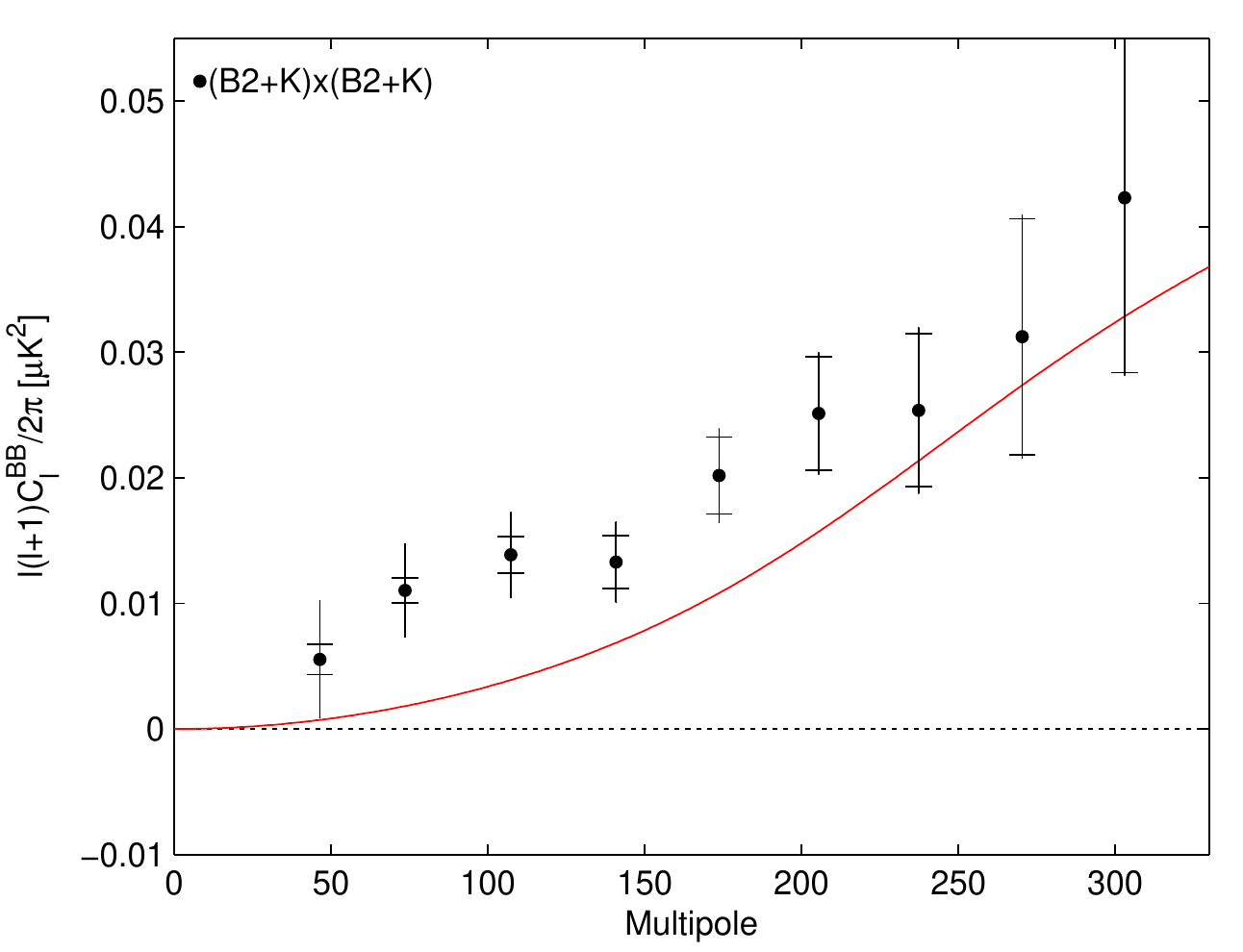}}
\end{center}
\caption{The $BB$ power spectrum of combined \biceptwo\ and
\keckarray\ maps.
The inner error bars are the standard deviation of the lensed-\lcdm+noise simulations,
while the outer error bars also contain excess power at low-$\ell$.}
\label{fig:b2kcomb}
\end{figure}

The final $BB$ spectrum is shown in Figure~\ref{fig:b2kcomb}
and is inconsistent with the lensed-\lcdm\ expectation
at $>6\sigma$ (for either band powers 1--5 or 1--9).
The lensed-\lcdm+noise error bars as plotted
are approximately a factor two smaller than those
of the previous \biceptwo\ only results---saturation on
the (small) sample variance of the lensing component is
occurring---the noise component is a factor 2.3 times
smaller.
All the spectra (including $TT$, $EE$ etc.) are available
for download at \url{http://bicepkeck.org/} together with
the ancillary data, noise information etc., required to
use them.

\section{Conclusions}
\label{sec:conc}

We have presented the \keckarray\ instrument
and the 2012--2013 (150~GHz) data set.
The instantaneous instrumental sensitivity of 9.5~\ukcmbrts is the best
reported to date.
The same area of sky as previously observed by \biceptwo\ was
mapped to a depth in $Q$ and $U$ of 74~nK$\,$deg (4.4 $\mu$K$\,$arcmin).
The resulting \keckarray\ power spectra are consistent with lensed-\lcdm\
except for an excess at degree angular scales in $BB$
which has a significance of $5.0\sigma$.
Extensive jackknife tests argue against a systematic origin for the
signal, and further statistical tests indicate that the maps and
spectra are consistent with the previous \biceptwo\ results.
Finally the two sets of maps are combined to
produce maps with noise of 57~nK$\,$deg (3.4~~$\mu$K$\,$arcmin)
over an effective area of 400 deg$^2$
for a survey weight of 250,000~$\mu$K$^{-2}$.
The final $BB$ spectrum is inconsistent with lensed-\lcdm\ at
a significance of $>6\sigma$.
The combined map results (for all spectra) are available for
download.
There does not appear to be any reason to consider the
\biceptwo\ results as more reliable than the \keckarray\
results or vice versa.
We therefore emphasize that we regard the combined results as the
best available data set at this time.

The origin of the excess power shown in
Figure~\ref{fig:b2kcomb}, and previously reported in the
\biceptwo\ Results Paper, has been extensively debated
in the literature~\citep{mortonson14,flauger14,fuskeland14}.
Recently, concrete information on the strength of polarized
dust emission at high galactic latitude has become available
in~\citet{planckiXXX}.
It appears that dust emission is a significant
contribution to the signal observed by
\biceptwo\ and the \keckarray.
Therefore, in an upcoming paper, the \biceptwo\ and
combined maps are cross correlated with
\planck\ maps of the same region to constrain the
dust contribution to the observed signal.

During the 2014 season two of the \keckarray\ receivers
operated at 95~GHz and a future analysis will use this data
to further constrain the dust contribution.
In the 2015 season \bicep3 will provide increased
sensitivity at 95~GHz and operation of \keckarray\ receivers
at 220~GHz is also planned.

\acknowledgements

The \keckarray\ project has been made possible through
support from the National Science Foundation under Grants
ANT-1145172 (Harvard), ANT-1145143 (Minnesota)
\& ANT-1145248 (Stanford), and from the Keck Foundation
(Caltech).
The development of antenna-coupled detector technology was supported
by the JPL Research and Technology Development Fund and Grants No.\
06-ARPA206-0040 and 10-SAT10-0017 from the NASA APRA and SAT programs.
The development and testing of focal planes were supported
by the Gordon and Betty Moore Foundation at Caltech.
Readout electronics were supported by a Canada Foundation
for Innovation grant to UBC.
The computations in this paper were run on the Odyssey cluster
supported by the FAS Science Division Research Computing Group at
Harvard University.
The analysis effort at Stanford and SLAC is partially supported by
the U.S. Department of Energy Office of Science.
We thank the staff of the U.S. Antarctic Program and in particular
the South Pole Station without whose help this research would not
have been possible.
Most special thanks go to our heroic winter-overs Robert Schwarz
and Steffen Richter.
We thank all those who have contributed past efforts to the \bicep--\keckarray\
series of experiments, including the \bicepone\ team.

\bibliographystyle{apj}
\bibliography{ms}

\begin{thebibliography}{}
\expandafter\ifx\csname natexlab\endcsname\relax\def\natexlab#1{#1}\fi

\bibitem[{Ade {et~al.}(2006)Ade, Pisano, Tucker, \& Weaver}]{ade06}
Ade, P. A.~R., Pisano, G., Tucker, C., \& Weaver, S. 2006, Proc. SPIE, 6275,
  62750U

\bibitem[{{Aikin} {et~al.}(2010){Aikin}, {Ade}, {Benton}, {Bock}, {Bonetti},
  {Brevik}, {Dowell}, {Duband}, {Filippini}, {Golwala}, {Halpern}, {Hristov},
  {Irwin}, {Kaufman}, {Keating}, {Kovac}, {Kuo}, {Lange}, {Netterfield},
  {Nguyen}, {Ogburn}, {Orlando}, {Pryke}, {Richter}, {Ruhl}, {Runyan},
  {Sheehy}, {Stokes}, {Sudiwala}, {Teply}, {Tolan}, {Turner}, {Wilson}, \&
  {Wong}}]{aikin10}
{Aikin}, R.~W., {Ade}, P.~A., {Benton}, S., {et~al.} 2010, in Society of
  Photo-Optical Instrumentation Engineers (SPIE) Conference Series, Vol. 7741,
  Society of Photo-Optical Instrumentation Engineers (SPIE) Conference Series,
  0

\bibitem[{{Barkats} {et~al.}(2005){Barkats}, {Bischoff}, {Farese},
  {Fitzpatrick}, {Gaier}, {Gundersen}, {Hedman}, {Hyatt}, {McMahon},
  {Samtleben}, {Staggs}, {Vanderlinde}, \& {Winstein}}]{barkats04}
{Barkats}, D., {Bischoff}, C., {Farese}, P., {et~al.} 2005, \apjl, 619, L127

\bibitem[{Battistelli {et~al.}(2008)}]{mce08}
Battistelli, E.~S., {et~al.} 2008, J. Low Temp. Phys., 151, 908

\bibitem[{{Bennett} {et~al.}(2013){Bennett}, {Larson}, {Weiland}, {Jarosik},
  {Hinshaw}, {Odegard}, {Smith}, {Hill}, {Gold}, {Halpern}, {Komatsu}, {Nolta},
  {Page}, {Spergel}, {Wollack}, {Dunkley}, {Kogut}, {Limon}, {Meyer}, {Tucker},
  \& {Wright}}]{bennett13}
{Bennett}, C.~L., {Larson}, D., {Weiland}, J.~L., {et~al.} 2013, \apjs, 208, 20

\bibitem[{{Benton} {et~al.}(2014){Benton}, {Ade}, {Amiri}, {Angil{\`e}},
  {Bock}, {Bond}, {Bryan}, {Chiang}, {Contaldi}, {Crill}, {Devlin}, {Dober},
  {Dor{\'e}}, {Farhang}, {Filippini}, {Fissel}, {Fraisse}, {Fukui}, {Galitzki},
  {Gambrel}, {Gandilo}, {Golwala}, {Gudmundsson}, {Halpern}, {Hasselfield},
  {Hilton}, {Holmes}, {Hristov}, {Irwin}, {Jones}, {Kermish}, {Klein},
  {Korotkov}, {Kuo}, {MacTavish}, {Mason}, {Matthews}, {Megerian}, {Moncelsi},
  {Morford}, {Mroczkowski}, {Nagy}, {Netterfield}, {Novak}, {Nutter},
  {O'Brient}, {Ogburn}, {Pascale}, {Poidevin}, {Rahlin}, {Reintsema}, {Ruhl},
  {Runyan}, {Savini}, {Scott}, {Shariff}, {Soler}, {Thomas}, {Trangsrud},
  {Truch}, {Tucker}, {Tucker}, {Tucker}, {Turner}, {Ward-Thompson}, {Weber},
  {Wiebe}, \& {Young}}]{benton14}
{Benton}, S.~J., {Ade}, P.~A., {Amiri}, M., {et~al.} 2014, in Society of
  Photo-Optical Instrumentation Engineers (SPIE) Conference Series, Vol. 9145,
  Society of Photo-Optical Instrumentation Engineers (SPIE) Conference Series,
  0

\bibitem[{{\textsc{Bicep1} Collaboration} {et~al.}(2014){\textsc{Bicep1}
  Collaboration}, {Barkats}, {Aikin}, {Bischoff}, {Buder}, {Kaufman},
  {Keating}, {Kovac}, {Su}, {Ade}, {Battle}, {Bierman}, {Bock}, {Chiang},
  {Dowell}, {Duband}, {Filippini}, {Hivon}, {Holzapfel}, {Hristov}, {Jones},
  {Kuo}, {Leitch}, {Mason}, {Matsumura}, {Nguyen}, {Ponthieu}, {Pryke},
  {Richter}, {Rocha}, {Sheehy}, {Kernasovskiy}, {Takahashi}, {Tolan}, \&
  {Yoon}}]{barkats14}
{\textsc{Bicep1} Collaboration}, {Barkats}, D., {Aikin}, R., {et~al.} 2014, The
  Astrophysical Journal, 783, 67

\bibitem[{{\textsc{Bicep2} Collaboration I}(2014)}]{b2respap14}
{\textsc{Bicep2} Collaboration I}. 2014, Phys. Rev. Lett., 112, 241101

\bibitem[{{\textsc{Bicep2} Collaboration II}(2014)}]{b2instpap14}
{\textsc{Bicep2} Collaboration II}. 2014, The Astrophysical Journal, 792, 62

\bibitem[{{\textsc{Bicep2} Collaboration III}(2015)}]{b2systpap14}
{\textsc{Bicep2} Collaboration III}. 2015, In prep.

\bibitem[{{\textsc{Bicep2}, \textit{Keck Array}, and \spider\
  Collaborations}(2015)}]{dets2014}
{\textsc{Bicep2}, \textit{Keck Array}, and \spider\ Collaborations}. 2015, In
  prep.

\bibitem[{{\textsc{Bicep2} / \textit{Keck Array} Collaborations
  IV}(2015)}]{b2beams14}
{\textsc{Bicep2} / \textit{Keck Array} Collaborations IV}. 2015, In prep.

\bibitem[{{Bischoff} {et~al.}(2008){Bischoff}, {Hyatt}, {McMahon}, {Nixon},
  {Samtleben}, {Smith}, {Vanderlinde}, {Barkats}, {Farese}, {Gaier},
  {Gundersen}, {Hedman}, {Staggs}, {Winstein}, \& {CAPMAP
  Collaboration}}]{bischoff08}
{Bischoff}, C., {Hyatt}, L., {McMahon}, J.~J., {et~al.} 2008, \apj, 684, 771

\bibitem[{{Brown} {et~al.}(2009){Brown}, {Ade}, {Bock}, {Bowden}, {Cahill},
  {Castro}, {Church}, {Culverhouse}, {Friedman}, {Ganga}, {Gear}, {Gupta},
  {Hinderks}, {Kovac}, {Lange}, {Leitch}, {Melhuish}, {Memari}, {Murphy},
  {Orlando}, {O'Sullivan}, {Piccirillo}, {Pryke}, {Rajguru}, {Rusholme},
  {Schwarz}, {Taylor}, {Thompson}, {Turner}, {Wu}, {Zemcov}, \& {QUa D
  Collaboration}}]{brown09}
{Brown}, M.~L., {Ade}, P., {Bock}, J., {et~al.} 2009, \apj, 705, 978

\bibitem[{{Caligiuri} \& {Kosowsky}(2014)}]{caligiuri14}
{Caligiuri}, J., \& {Kosowsky}, A. 2014, Physical Review Letters, 112, 191302

\bibitem[{{Chiang} {et~al.}(2010){Chiang}, {Ade}, {Barkats}, {Battle},
  {Bierman}, {Bock}, {Dowell}, {Duband}, {Hivon}, {Holzapfel}, {Hristov},
  {Jones}, {Keating}, {Kovac}, {Kuo}, {Lange}, {Leitch}, {Mason}, {Matsumura},
  {Nguyen}, {Ponthieu}, {Pryke}, {Richter}, {Rocha}, {Sheehy}, {Takahashi},
  {Tolan}, \& {Yoon}}]{chiang10}
{Chiang}, H.~C., {Ade}, P.~A.~R., {Barkats}, D., {et~al.} 2010, \apj, 711, 1123

\bibitem[{{Crites} {et~al.}(2014){Crites}, {Henning}, {Ade}, {Aird},
  {Austermann}, {Beall}, {Bender}, {Benson}, {Bleem}, {Carstrom}, {Chang},
  {Chiang}, {Cho}, {Citron}, {Crawford}, {De Haan}, {Dobbs}, {Everett},
  {Gallicchio}, {Gao}, {George}, {Gilbert}, {Halverson}, {Hanson},
  {Harrington}, {Hilton}, {Holder}, {Holzapfel}, {Hoover}, {Hou}, {Hrubes},
  {Huang}, {Hubmayr}, {Irwin}, {Keisler}, {Knox}, {Lee}, {Leitch}, {Li},
  {Liang}, {Luong-Van}, {McMahon}, {Mehl}, {Meyer}, {Mocanu}, {Montroy},
  {Natoli}, {Nibarger}, {Novosad}, {Padin}, {Pryke}, {Reichardt}, {Ruhl},
  {Saliwanchik}, {Sayre}, {Schaffer}, {Smecher}, {Stark}, {Story}, {Tucker},
  {Vanderlinde}, {Vieira}, {Wang}, {Whitehorn}, {Yefremenko}, \&
  {Zahn}}]{crites14}
{Crites}, A.~T., {Henning}, J.~W., {Ade}, P.~A.~R., {et~al.} 2014, ArXiv
  e-prints, arXiv:1411.1042

\bibitem[{{de Korte} {et~al.}(2003){de Korte}, {Beyer}, {Deiker}, {Hilton},
  {Irwin}, {Macintosh}, {Nam}, {Reintsema}, {Vale}, \& {Huber}}]{dekorte03}
{de Korte}, P.~A.~J., {Beyer}, J., {Deiker}, S., {et~al.} 2003, Review of
  Scientific Instruments, 74, 3807

\bibitem[{{Dodelson}(2014)}]{dodelson14}
{Dodelson}, S. 2014, Physical Review Letters, 112, 191301

\bibitem[{Duband \& Collaudin(1999)}]{duband99}
Duband, L., \& Collaudin, B. 1999, Cryogenics, 39, 659

\bibitem[{{Flauger} {et~al.}(2014){Flauger}, {Hill}, \& {Spergel}}]{flauger14}
{Flauger}, R., {Hill}, J.~C., \& {Spergel}, D.~N. 2014, arXiv:1405.7351v1

\bibitem[{{Fuskeland} {et~al.}(2014){Fuskeland}, {Wehus}, {Eriksen}, \&
  {N{\ae}ss}}]{fuskeland14}
{Fuskeland}, U., {Wehus}, I.~K., {Eriksen}, H.~K., \& {N{\ae}ss}, S.~K. 2014,
  \apj, 790, 104

\bibitem[{{Galli} {et~al.}(2014){Galli}, {Benabed}, {Bouchet}, {Cardoso},
  {Elsner}, {Hivon}, {Mangilli}, {Prunet}, \& {Wandelt}}]{galli14}
{Galli}, S., {Benabed}, K., {Bouchet}, F., {et~al.} 2014, Phys. Rev. D, 90,
  063504

\bibitem[{{G{\'o}rski} {et~al.}(2005){G{\'o}rski}, {Hivon}, {Banday},
  {Wandelt}, {Hansen}, {Reinecke}, \& {Bartelmann}}]{healpix}
{G{\'o}rski}, K.~M., {Hivon}, E., {Banday}, A.~J., {et~al.} 2005, \apj, 622,
  759

\bibitem[{{Hanson} {et~al.}(2013){Hanson}, {Hoover}, {Crites}, {Ade}, {Aird},
  {Austermann}, {Beall}, {Bender}, {Benson}, {Bleem}, {Bock}, {Carlstrom},
  {Chang}, {Chiang}, {Cho}, {Conley}, {Crawford}, {de Haan}, {Dobbs},
  {Everett}, {Gallicchio}, {Gao}, {George}, {Halverson}, {Harrington},
  {Henning}, {Hilton}, {Holder}, {Holzapfel}, {Hrubes}, {Huang}, {Hubmayr},
  {Irwin}, {Keisler}, {Knox}, {Lee}, {Leitch}, {Li}, {Liang}, {Luong-Van},
  {Marsden}, {McMahon}, {Mehl}, {Meyer}, {Mocanu}, {Montroy}, {Natoli},
  {Nibarger}, {Novosad}, {Padin}, {Pryke}, {Reichardt}, {Ruhl}, {Saliwanchik},
  {Sayre}, {Schaffer}, {Schulz}, {Smecher}, {Stark}, {Story}, {Tucker},
  {Vanderlinde}, {Vieira}, {Viero}, {Wang}, {Yefremenko}, {Zahn}, \&
  {Zemcov}}]{hanson13}
{Hanson}, D., {Hoover}, S., {Crites}, A., {et~al.} 2013, Physical Review
  Letters, 111, 141301

\bibitem[{{Hinderks} {et~al.}(2009){Hinderks}, {Ade}, {Bock}, {Bowden},
  {Brown}, {Cahill}, {Carlstrom}, {Castro}, {Church}, {Culverhouse},
  {Friedman}, {Ganga}, {Gear}, {Gupta}, {Harris}, {Haynes}, {Keating}, {Kovac},
  {Kirby}, {Lange}, {Leitch}, {Mallie}, {Melhuish}, {Memari}, {Murphy},
  {Orlando}, {Schwarz}, {Sullivan}, {Piccirillo}, {Pryke}, {Rajguru},
  {Rusholme}, {Taylor}, {Thompson}, {Tucker}, {Turner}, {Wu}, \&
  {Zemcov}}]{hinderks09}
{Hinderks}, J.~R., {Ade}, P., {Bock}, J., {et~al.} 2009, \apj, 692, 1221

\bibitem[{{Hivon} {et~al.}(2002){Hivon}, {G{\'o}rski}, {Netterfield}, {Crill},
  {Prunet}, \& {Hansen}}]{hivon02}
{Hivon}, E., {G{\'o}rski}, K.~M., {Netterfield}, C.~B., {et~al.} 2002, \apj,
  567, 2

\bibitem[{Irwin \& Hilton(2005)}]{hiltonirwin2005}
Irwin, K., \& Hilton, G. 2005, in Topics in Applied Physics, Vol.~99, Cryogenic
  Particle Detection, ed. C.~Enss (Springer Berlin Heidelberg), 63--150

\bibitem[{{Kamionkowski} {et~al.}(1997){Kamionkowski}, {Kosowsky}, \&
  {Stebbins}}]{kamionkowski97}
{Kamionkowski}, M., {Kosowsky}, A., \& {Stebbins}, A. 1997, Physical Review
  Letters, 78, 2058

\bibitem[{{Karkare} {et~al.}(2014){Karkare}, {Ade}, {Ahmed}, {Aikin},
  {Alexander}, {Amiri}, {Barkats}, {Benton}, {Bischoff}, {Bock}, {Bonetti},
  {Brevik}, {Buder}, {Bullock}, {Burger}, {Connors}, {Crill}, {Davis},
  {Dowell}, {Duband}, {Filippini}, {Fliescher}, {Golwala}, {Gordon}, {Grayson},
  {Halpern}, {Hasselfield}, {Hildebrandt}, {Hilton}, {Hristov}, {Hui}, {Irwin},
  {Kang}, {Karpel}, {Kefeli}, {Kernasovskiy}, {Kovac}, {Kuo}, {Leitch},
  {Lueker}, {Mason}, {Megerian}, {Netterfield}, {Nguyen}, {O'Brient}, {Ogburn},
  {Pryke}, {Reintsema}, {Richter}, {Schwarz}, {Sheehy}, {Staniszewski},
  {Sudiwala}, {Teply}, {Thompson}, {Tolan}, {Turner}, {Vieregg}, {Weber},
  {Wong}, {Wu}, \& {Yoon}}]{karkare14}
{Karkare}, K.~S., {Ade}, P.~A.~R., {Ahmed}, Z., {et~al.} 2014, in Society of
  Photo-Optical Instrumentation Engineers (SPIE) Conference Series, Vol. 9153,
  Society of Photo-Optical Instrumentation Engineers (SPIE) Conference Series,
  3

\bibitem[{{Kaufman} {et~al.}(2014){Kaufman}, {Miller}, {Shimon}, {Barkats},
  {Bischoff}, {Buder}, {Keating}, {Kovac}, {Ade}, {Aikin}, {Battle}, {Bierman},
  {Bock}, {Chiang}, {Dowell}, {Duband}, {Filippini}, {Hivon}, {Holzapfel},
  {Hristov}, {Jones}, {Kernasovskiy}, {Kuo}, {Leitch}, {Mason}, {Matsumura},
  {Nguyen}, {Ponthieu}, {Pryke}, {Richter}, {Rocha}, {Sheehy}, {Su},
  {Takahashi}, {Tolan}, \& {Yoon}}]{kaufman14}
{Kaufman}, J.~P., {Miller}, N.~J., {Shimon}, M., {et~al.} 2014, \prd, 89,
  062006

\bibitem[{{Kernasovskiy} {et~al.}(2012)}]{kernasovskiy12}
{Kernasovskiy}, S., {et~al.} 2012, in Society of Photo-Optical Instrumentation
  Engineers (SPIE) Conference Series, Vol. 8452, Society of Photo-Optical
  Instrumentation Engineers (SPIE) Conference Series

\bibitem[{{Kovac} {et~al.}(2002){Kovac}, {Leitch}, {Pryke}, {Carlstrom},
  {Halverson}, \& {Holzapfel}}]{kovac02}
{Kovac}, J.~M., {Leitch}, E.~M., {Pryke}, C., {et~al.} 2002, Nature, 420, 772

\bibitem[{{Leitch} {et~al.}(2002){Leitch}, {Kovac}, {Pryke}, {Carlstrom},
  {Halverson}, {Holzapfel}, {Dragovan}, {Reddall}, \& {Sandberg}}]{leitch02}
{Leitch}, E.~M., {Kovac}, J.~M., {Pryke}, C., {et~al.} 2002, \nat, 420, 763

\bibitem[{{Montroy} {et~al.}(2006){Montroy}, {Ade}, {Bock}, {Bond}, {Borrill},
  {Boscaleri}, {Cabella}, {Contaldi}, {Crill}, {de Bernardis}, {De Gasperis},
  {de Oliveira-Costa}, {De Troia}, {di Stefano}, {Hivon}, {Jaffe}, {Kisner},
  {Jones}, {Lange}, {Masi}, {Mauskopf}, {MacTavish}, {Melchiorri}, {Natoli},
  {Netterfield}, {Pascale}, {Piacentini}, {Pogosyan}, {Polenta}, {Prunet},
  {Ricciardi}, {Romeo}, {Ruhl}, {Santini}, {Tegmark}, {Veneziani}, \&
  {Vittorio}}]{montroy06}
{Montroy}, T.~E., {Ade}, P.~A.~R., {Bock}, J.~J., {et~al.} 2006, \apj, 647, 813

\bibitem[{{Mortonson} \& {Seljak}(2014)}]{mortonson14}
{Mortonson}, M.~J., \& {Seljak}, U. 2014, arXiv:1405.5857

\bibitem[{{Naess} {et~al.}(2014){Naess}, {Hasselfield}, {McMahon}, {Niemack},
  {Addison}, {Ade}, {Allison}, {Amiri}, {Baker}, {Battaglia}, {Beall}, {de
  Bernardis}, {Bond}, {Britton}, {Calabrese}, {Cho}, {Coughlin}, {Crichton},
  {Das}, {Datta}, {Devlin}, {Dicker}, {Dunkley}, {D{\"u}nner}, {Fowler}, {Fox},
  {Gallardo}, {Grace}, {Gralla}, {Hajian}, {Halpern}, {Henderson}, {Hill},
  {Hilton}, {Hilton}, {Hincks}, {Hlozek}, {Ho}, {Hubmayr}, {Huffenberger},
  {Hughes}, {Infante}, {Irwin}, {Jackson}, {Klein}, {Koopman}, {Kosowsky},
  {Li}, {Louis}, {Lungu}, {Madhavacheril}, {Marriage}, {Maurin}, {Menanteau},
  {Moodley}, {Munson}, {Newburgh}, {Nibarger}, {Nolta}, {Page}, {Pappas},
  {Partridge}, {Rojas}, {Schmitt}, {Sehgal}, {Sherwin}, {Sievers}, {Simon},
  {Spergel}, {Staggs}, {Switzer}, {Thornton}, {Trac}, {Tucker}, {Van Engelen},
  {Ward}, \& {Wollack}}]{naess14}
{Naess}, S., {Hasselfield}, M., {McMahon}, J., {et~al.} 2014, \jcap, 10, 7

\bibitem[{{Page} {et~al.}(2007){Page}, {Hinshaw}, {Komatsu}, {Nolta},
  {Spergel}, {Bennett}, {Barnes}, {Bean}, {Dor{\'e}}, {Dunkley}, {Halpern},
  {Hill}, {Jarosik}, {Kogut}, {Limon}, {Meyer}, {Odegard}, {Peiris}, {Tucker},
  {Verde}, {Weiland}, {Wollack}, \& {Wright}}]{page07}
{Page}, L., {Hinshaw}, G., {Komatsu}, E., {et~al.} 2007, \apjs, 170, 335

\bibitem[{{Planck Collaboration Int.\ XXX}(2014)}]{planckiXXX}
{Planck Collaboration Int.\ XXX}. 2014, \aap, arXiv:1409.5738

\bibitem[{{Planck Collaboration results I}(2014)}]{planckI}
{Planck Collaboration results I}. 2014, \aap, 571, A1

\bibitem[{{\textsc{Polarbear} Collaboration}
  {et~al.}(2014{\natexlab{a}}){\textsc{Polarbear} Collaboration}, {Ade},
  {Akiba}, {Anthony}, {Arnold}, {Atlas}, {Barron}, {Boettger}, {Borrill},
  {Chapman}, {Chinone}, {Dobbs}, {Elleflot}, {Errard}, {Fabbian}, {Feng},
  {Flanigan}, {Gilbert}, {Grainger}, {Halverson}, {Hasegawa}, {Hattori},
  {Hazumi}, {Holzapfel}, {Hori}, {Howard}, {Hyland}, {Inoue}, {Jaehnig},
  {Jaffe}, {Keating}, {Kermish}, {Keskitalo}, {Kisner}, {Le Jeune}, {Lee},
  {Leitch}, {Linder}, {Lungu}, {Matsuda}, {Matsumura}, {Meng}, {Miller},
  {Morii}, {Moyerman}, {Myers}, {Navaroli}, {Nishino}, {Paar}, {Peloton},
  {Poletti}, {Quealy}, {Rebeiz}, {Reichardt}, {Richards}, {Ross}, {Schanning},
  {Schenck}, {Sherwin}, {Shimizu}, {Shimmin}, {Shimon}, {Siritanasak},
  {Smecher}, {Spieler}, {Stebor}, {Steinbach}, {Stompor}, {Suzuki}, {Takakura},
  {Tomaru}, {Wilson}, {Yadav}, \& {Zahn}}]{polarbear14}
{\textsc{Polarbear} Collaboration}, {Ade}, P.~A.~R., {Akiba}, Y., {et~al.}
  2014{\natexlab{a}}, \apj, 794, 171

\bibitem[{{\textsc{Polarbear} Collaboration}
  {et~al.}(2014{\natexlab{b}}){\textsc{Polarbear} Collaboration}, {Ade},
  {Akiba}, {Anthony}, {Arnold}, {Barron}, {Boettger}, {Borrill}, {Borys},
  {Chapman}, {Chinone}, {Dobbs}, {Elleflot}, {Errard}, {Fabbian}, {Feng},
  {Flanigan}, {Gilbert}, {Grainger}, {Halverson}, {Hasegawa}, {Hattori},
  {Hazumi}, {Holzapfel}, {Hori}, {Howard}, {Hyland}, {Inoue}, {Jaehnig},
  {Jaffe}, {Keating}, {Kermish}, {Keskitalo}, {Kisner}, {Le Jeune}, {Lee},
  {Linder}, {Lungu}, {Matsuda}, {Matsumura}, {Meng}, {Miller}, {Morii},
  {Moyerman}, {Myers}, {Navaroli}, {Nishino}, {Paar}, {Peloton}, {Quealy},
  {Rebeiz}, {Reichardt}, {Richards}, {Ross}, {Rotermund}, {Schanning},
  {Schenck}, {Sherwin}, {Shimizu}, {Shimmin}, {Shimon}, {Siritanasak},
  {Smecher}, {Spieler}, {Stebor}, {Steinbach}, {Stompor}, {Suzuki}, {Takakura},
  {Tikhomirov}, {Tomaru}, {Wilson}, {Yadav}, \& {Zahn}}]{polarbear6645}
---. 2014{\natexlab{b}}, Physical Review Letters, 112, 131302

\bibitem[{{\textsc{Polarbear} Collaboration}
  {et~al.}(2014{\natexlab{c}}){\textsc{Polarbear} Collaboration}, {Ade},
  {Akiba}, {Anthony}, {Arnold}, {Barron}, {Boettger}, {Borrill}, {Chapman},
  {Chinone}, {Dobbs}, {Elleflot}, {Errard}, {Fabbian}, {Feng}, {Flanigan},
  {Gilbert}, {Grainger}, {Halverson}, {Hasegawa}, {Hattori}, {Hazumi},
  {Holzapfel}, {Hori}, {Howard}, {Hyland}, {Inoue}, {Jaehnig}, {Jaffe},
  {Keating}, {Kermish}, {Keskitalo}, {Kisner}, {Le Jeune}, {Lee}, {Linder},
  {Lungu}, {Matsuda}, {Matsumura}, {Meng}, {Miller}, {Morii}, {Moyerman},
  {Myers}, {Navaroli}, {Nishino}, {Paar}, {Peloton}, {Quealy}, {Rebeiz},
  {Reichardt}, {Richards}, {Ross}, {Schanning}, {Schenck}, {Sherwin},
  {Shimizu}, {Shimmin}, {Shimon}, {Siritanasak}, {Smecher}, {Spieler},
  {Stebor}, {Steinbach}, {Stompor}, {Suzuki}, {Takakura}, {Tomaru}, {Wilson},
  {Yadav}, \& {Zahn}}]{polarbear6646}
---. 2014{\natexlab{c}}, Physical Review Letters, 113, 021301

\bibitem[{{Polnarev}(1985)}]{polnarev85}
{Polnarev}, A.~G. 1985, \sovast, 29, 607

\bibitem[{{Pryke} {et~al.}(2009){Pryke}, {Ade}, {Bock}, {Bowden}, {Brown},
  {Cahill}, {Castro}, {Church}, {Culverhouse}, {Friedman}, {Ganga}, {Gear},
  {Gupta}, {Hinderks}, {Kovac}, {Lange}, {Leitch}, {Melhuish}, {Memari},
  {Murphy}, {Orlando}, {Schwarz}, {Sullivan}, {Piccirillo}, {Rajguru},
  {Rusholme}, {Taylor}, {Thompson}, {Turner}, {Wu}, \& {Zemcov}}]{pryke09}
{Pryke}, C., {Ade}, P., {Bock}, J., {et~al.} 2009, \apj, 692, 1247

\bibitem[{{\textsc{Quiet} Collaboration} {et~al.}(2011){\textsc{Quiet}
  Collaboration}, {Bischoff}, {Brizius}, {Buder}, {Chinone}, {Cleary},
  {Dumoulin}, {Kusaka}, {Monsalve}, {N{\ae}ss}, {Newburgh}, {Reeves}, {Smith},
  {Wehus}, {Zuntz}, {Zwart}, {Bronfman}, {Bustos}, {Church}, {Dickinson},
  {Eriksen}, {Ferreira}, {Gaier}, {Gundersen}, {Hasegawa}, {Hazumi},
  {Huffenberger}, {Jones}, {Kangaslahti}, {Kapner}, {Lawrence}, {Limon}, {May},
  {McMahon}, {Miller}, {Nguyen}, {Nixon}, {Pearson}, {Piccirillo}, {Radford},
  {Readhead}, {Richards}, {Samtleben}, {Seiffert}, {Shepherd}, {Staggs},
  {Tajima}, {Thompson}, {Vanderlinde}, {Williamson}, \& {Winstein}}]{quiet11}
{\textsc{Quiet} Collaboration}, {Bischoff}, C., {Brizius}, A., {et~al.} 2011,
  \apj, 741, 111

\bibitem[{{\textsc{Quiet} Collaboration} {et~al.}(2012){\textsc{Quiet}
  Collaboration}, {Araujo}, {Bischoff}, {Brizius}, {Buder}, {Chinone},
  {Cleary}, {Dumoulin}, {Kusaka}, {Monsalve}, {N{\ae}ss}, {Newburgh}, {Reeves},
  {Wehus}, {Zwart}, {Bronfman}, {Bustos}, {Church}, {Dickinson}, {Eriksen},
  {Gaier}, {Gundersen}, {Hasegawa}, {Hazumi}, {Huffenberger}, {Ishidoshiro},
  {Jones}, {Kangaslahti}, {Kapner}, {Kubik}, {Lawrence}, {Limon}, {McMahon},
  {Miller}, {Nagai}, {Nguyen}, {Nixon}, {Pearson}, {Piccirillo}, {Radford},
  {Readhead}, {Richards}, {Samtleben}, {Seiffert}, {Shepherd}, {Smith},
  {Staggs}, {Tajima}, {Thompson}, {Vanderlinde}, \& {Williamson}}]{quiet12}
{\textsc{Quiet} Collaboration}, {Araujo}, D., {Bischoff}, C., {et~al.} 2012,
  \apj, 760, 145

\bibitem[{{Readhead} {et~al.}(2004){Readhead}, {Myers}, {Pearson}, {Sievers},
  {Mason}, {Contaldi}, {Bond}, {Bustos}, {Altamirano}, {Achermann}, {Bronfman},
  {Carlstrom}, {Cartwright}, {Casassus}, {Dickinson}, {Holzapfel}, {Kovac},
  {Leitch}, {May}, {Padin}, {Pogosyan}, {Pospieszalski}, {Pryke}, {Reeves},
  {Shepherd}, \& {Torres}}]{readhead04}
{Readhead}, A.~C.~S., {Myers}, S.~T., {Pearson}, T.~J., {et~al.} 2004, Science,
  306, 836

\bibitem[{{Rocha} {et~al.}(2004){Rocha}, {Trotta}, {Martins}, {Melchiorri},
  {Avelino}, {Bean}, \& {Viana}}]{rocha04}
{Rocha}, G., {Trotta}, R., {Martins}, C.~J.~A.~P., {et~al.} 2004, \mnras, 352,
  20

\bibitem[{{Seljak}(1997)}]{seljak97b}
{Seljak}, U. 1997, \apj, 482, 6

\bibitem[{{Seljak} \& {Zaldarriaga}(1997)}]{seljak97a}
{Seljak}, U., \& {Zaldarriaga}, M. 1997, Physical Review Letters, 78, 2054

\bibitem[{{Sheehy} {et~al.}(2010)}]{sheehy10}
{Sheehy}, C.~D., {et~al.} 2010, Millimeter, Submillimeter, and Far-Infrared
  Detectors and Instrumentation for Astronomy V, 7741, 77411G

\bibitem[{{Sievers} {et~al.}(2007){Sievers}, {Achermann}, {Bond}, {Bronfman},
  {Bustos}, {Contaldi}, {Dickinson}, {Ferreira}, {Jones}, {Lewis}, {Mason},
  {May}, {Myers}, {Oyarce}, {Padin}, {Pearson}, {Pospieszalski}, {Readhead},
  {Reeves}, {Taylor}, \& {Torres}}]{sievers07}
{Sievers}, J.~L., {Achermann}, C., {Bond}, J.~R., {et~al.} 2007, \apj, 660, 976

\bibitem[{Stiehl {et~al.}(2011)Stiehl, Cho, Hilton, Irwin, Mates, Reintsema, \&
  Zink}]{stiehl11}
Stiehl, G., Cho, H.-M., Hilton, G., {et~al.} 2011, Applied Superconductivity,
  IEEE Transactions on, 21, 298

\bibitem[{{Story} {et~al.}(2012){Story}, {Leitch}, {Ade}, {Aird}, {Austermann},
  {Beall}, {Becker}, {Bender}, {Benson}, {Bleem}, {Britton}, {Carlstrom},
  {Chang}, {Chiang}, {Cho}, {Crawford}, {Crites}, {Datesman}, {de Haan},
  {Dobbs}, {et~al.}}]{story12}
{Story}, K., {Leitch}, E., {Ade}, P., {et~al.} 2012, Proc.\ SPIE Int.\ Soc.\
  Opt.\ Eng., 8451, 84510T

\bibitem[{{Takahashi} {et~al.}(2010){Takahashi}, {Ade}, {Barkats}, {Battle},
  {Bierman}, {Bock}, {Chiang}, {Dowell}, {Duband}, {Hivon}, {Holzapfel},
  {Hristov}, {Jones}, {Keating}, {Kovac}, {Kuo}, {Lange}, {Leitch}, {Mason},
  {Matsumura}, {Nguyen}, {Ponthieu}, {Pryke}, {Richter}, {Rocha}, \&
  {Yoon}}]{takahashi10}
{Takahashi}, Y.~D., {Ade}, P.~A.~R., {Barkats}, D., {et~al.} 2010, \apj, 711,
  1141

\bibitem[{{Wu} {et~al.}(2007){Wu}, {Zuntz}, {Abroe}, {Ade}, {Bock}, {Borrill},
  {Collins}, {Hanany}, {Jaffe}, {Johnson}, {Jones}, {Lee}, {Matsumura},
  {Rabii}, {Renbarger}, {Richards}, {Smoot}, {Stompor}, {Tran}, \&
  {Winant}}]{wu07}
{Wu}, J.~H.~P., {Zuntz}, J., {Abroe}, M.~E., {et~al.} 2007, \apj, 665, 55

\bibitem[{{Zaldarriaga} \& {Seljak}(1998)}]{zaldarriaga98}
{Zaldarriaga}, M., \& {Seljak}, U. 1998, \prd, 58, 023003

\end{thebibliography}

\end{document}